\documentclass[12pt]{article}

\usepackage{amstext}
\usepackage{graphicx}
\usepackage{amsmath,amssymb}
\usepackage{epsfig} 
\usepackage[bookmarksnumbered]{hyperref}

\def\square{\kern1pt\vbox{\hrule height 1.2pt\hbox{\vrule width 1.2pt\hskip 3pt
   \vbox{\vskip 6pt}\hskip 3pt\vrule width 0.6pt}\hrule height 0.6pt}\kern1pt}
\def\gtwid{\mathrel{\raise.3ex\hbox{$>$\kern-.75em\lower1ex\hbox{$\sim$}}}}
\def\ltwid{\mathrel{\raise.3ex\hbox{$<$\kern-.75em\lower1ex\hbox{$\sim$}}}}
\def\square{\kern1pt\vbox{\hrule height 1.2pt\hbox{\vrule width 1.2pt\hskip 3pt
   \vbox{\vskip 6pt}\hskip 3pt\vrule width 0.6pt}\hrule height 0.6pt}\kern1pt}

\begin{document}

\begin{titlepage}

\begin{flushright}
UFIFT-QG-14-07
\end{flushright}

\vskip 2cm

\begin{center}
{\bf One-loop quantum electrodynamic correction to the gravitational potentials on de Sitter spacetime}
\end{center}

\vskip 2cm

\begin{center}
C. L. Wang$^{*}$ and R. P. Woodard$^{\dagger}$
\end{center}

\begin{center}
\it{Department of Physics, University of Florida, Gainesville, Florida 32611, USA}
\end{center}

\vspace{1cm}

\begin{center}
ABSTRACT
\end{center}
We compute the one-loop photon contribution to the graviton self-energy 
on a de Sitter background and use it to solve the linearized Einstein 
equation for a point mass. Our results show that a comoving observer 
sees a logarithmic spatial running Newton's constant. Equivalently, a 
static observer reports a secular suppression of the Newtonian potential.

\begin{flushleft}
\end{flushleft}

\vskip 2cm

\begin{flushleft}
$^{*}$ e-mail: clwang@ufl.edu \\
$^{\dagger}$ e-mail: woodard@phys.ufl.edu
\end{flushleft}

\end{titlepage}

\section{INTRODUCTION}\label{intro}
From Newton's law of universal gravitation to Einstein's general relativity, 
our understanding of gravity has developed profoundly. However, efforts to
incorporate quantum mechanics into gravity have been unsuccessful, at least 
partly due to the fact that quantum gravitational effects are unobservably 
weak at low energies \cite{Woodard:2009ns}. This unsatisfactory situation 
has been transformed by the accumulation of large cosmological data sets on
primordial perturbations which are predicted by the theory of inflation to
arise from the gravitational response to quantum fluctuations of matter and 
from quantum fluctuations in gravitational radiation \cite{Starobinsky:1979ty,
Mukhanov:1981xt,Hawking:1982cz,Guth:1982ec,Starobinsky:1982ee,Bardeen:1983qw,
Mukhanov:1985rz,Mukhanov:1990me,Liddle:1993fq,Lidsey:1995np,Krauss:2013pha}. 
These quantum gravitational effects are hugely enhanced during primordial 
inflation and then fossilize so that they can survive to much later times
\cite{Woodard:2014jba}.

Electromagnetism provided humanity's first and simplest example of a 
relativistic unified field theory. It is therefore natural to wonder how
electromagnetism affects quantum gravity during inflation. Of course, Einstein plus Maxwell is not perturbatively renormalizable, even at one-loop 
order \cite{Deser:1974zzd,Deser:1974cz}, and that does mean we must accept
some limitations on what can be studied with it. We use Einstein plus Maxwell 
in the standard sense of effective field theory, absorbing ultraviolet 
divergences order-by-order with Bogoliubov, Parsiuk, Hepp ,and Zimmermann (BPHZ) counterterms \cite{BPHZ}. Once this is done, the renormalized
results from loops of massless gravitons and photons engender nonlocal and 
ultraviolet finite contributions to the effective action which are unique 
predictions of the theory that cannot be changed by its still unknown 
ultraviolet completion \cite{Donoghue:1993eb,Donoghue:1994dn}. A famous 
example is the computation by Bjerrum-Bohr of the one graviton loop 
correction to the Coulomb potential on a flat space background \cite{BjerrumBohr:2002sx}. We work in the same theoretical context but on a 
de Sitter background, and with the slightly different goal of inferring the
one photon loop corrections to the Newtonian potential.

One way of studying quantum weak-field corrections to a classical theory is 
to first compute the corresponding one-particle-irreducible (1PI) two-point 
function. This is then used to quantum correct the linearized effective field 
equations. For our problem we must compute the one photon loop contribution 
to the graviton self-energy $-i\bigl[{}^{\mu\nu}\Sigma^{\rho\sigma}\bigr](x;x')$.
We can then infer quantum corrections to graviton mode functions, or to the 
gravitational response to sources, from the linearized effective field equation
for the graviton field $h_{\mu\nu}(x)$,
\begin{equation}
\mathcal{D}^{\mu\nu\rho\sigma}\kappa h_{\rho\sigma}(x) - \int d^4x' \bigl[{}^{\mu\nu}\Sigma^{\rho\sigma}\bigr](x;x') \kappa h_{\rho\sigma}(x') = \mathcal{T}_{\text{lin}}^{\mu\nu}(x) \equiv -\kappa\frac{\delta 
S_{\text{matter}}}{\delta h_{\mu\nu}(x)}\biggl\vert_{h=0} \; , \label{lineffeq}
\end{equation}
Here, $\mathcal{D}^{\mu\nu\rho\sigma}$ is the Lichnerowicz operator in a de Sitter 
background, $\kappa^2 \equiv 16\pi G$ is the loop counting parameter of quantum 
gravity, and $\mathcal{T}_{\text{lin}}^{\mu\nu}(x)$ is the linearized stress 
tensor density of whatever source is desired. In our case this source is that 
of a static point mass. The same framework (with zero source) has been employed 
to investigate corrections to graviton mode functions from massless, minimally 
coupled (MMC) scalars \cite{Park:2010pj, Park:2011kg}. There have also been 
studies of the effect of a loop of massless, conformally invariant scalars on 
the graviton mode function \cite{Frob:2012ui,Frob:2014cza} and on the background 
geometry \cite{Frob:2013ht}.

This paper contains six sections, of which the first is this introduction. 
Section \ref{FRs} gives those of the Feynman rules for Maxwell plus Einstein which 
are needed for our computation, as well as some facts about the background 
geometry and the way we represent the tensor structure of the graviton self-energy.
In section \ref{GravSE}, we derive the primitive, dimensionally regulated result 
for the one photon loop contribution to the graviton self-energy. Section 
\ref{Renorm} introduces appropriate BPHZ counterterms for this theory and gives 
the result of renormalization. In section \ref{SolEffeqn}, we convert our in-out 
graviton self-energy to the retarded one of the Schwinger-Keldysh formalism, and we
use it to solve (\ref{lineffeq}) for a stationary point mass. In section 
\ref{discuss}, we propose a physical interpretation of our result.

\section{FEYNMAN RULES}\label{FRs}

\subsection{Preliminary clarifications}\label{prelim}
This work is based on quantum gravitational perturbation theory in de Sitter space which is known to be the unique maximally symmetric solution to the Einstein field equation with a positive cosmological constant $\Lambda$. But we will not work on the full D-dimensional de Sitter manifold rather on a submanifold known as a ``cosmological patch'' which is homogeneous, isotropic, and spatially flat. The invariant line element for this geometry is
\begin{equation}
ds^2 = -dt^2 + a^2(t) d\vec{x} \cdot d\vec{x} \; , \label{CosMetric}
\end{equation}
where $a(t)$ is the scale factor measuring the expansion of the Universe. The Hubble parameter is defined as $H(t)\equiv\dot{a}(t)/a(t)$. It is convenient to use open conformal coordinates $x^{\mu}=\left(\eta,x^i\right)$ with
\begin{equation}
-\infty < \eta <0 \;\; ,\;\; -\infty < x^i < +\infty \;\; ,\;\; \text{for} \;\;\; i=1,2,\cdots,D-1\; ,
\end{equation}
where the conformal time coordinate $\eta$ is related to the normal time coordinate $t$ through the relation, $dt\equiv ad\eta$. The metric in open conformal coordinates is conformally flat,
\begin{equation}
ds^2 = a^2(\eta)\left(-d\eta^2 + d\vec{x} \cdot d\vec{x}\right)=a^2(\eta)\eta_{\mu\nu}dx^{\mu}dx^{\nu} \; . \label{BgMetric}
\end{equation}
The Hubble parameter $H$ is constant in de Sitter spacetime, and the scale factor in terms of the conformal time coordinate is $a(\eta)=-\frac{1}{H\eta}$. We then define the graviton field $h_{\mu\nu}(x)$ by conformally transforming the full metric $g_{\mu\nu}(x)$ and then subtracting off the background,
\begin{equation}
g_{\mu\nu}(x) \equiv a^2(\eta)\Bigl[\eta_{\mu\nu}+\kappa h_{\mu\nu}(x)\Bigr]
\equiv a^2(\eta)\widetilde{g}_{\mu\nu}(x) \; , \label{FullMetric}
\end{equation}
where $\eta_{\mu\nu}$ is the D-dimensional Minkowski metric with a spacelike signature and $h_{\mu\nu}$ is the graviton field whose indices are raised and lowered with the Minkowski metric. And we will make great use of the de Sitter invariant length function $y(x;x')$ while representing our propagators on de Sitter spacetime,
\begin{equation}
y(x;x')\equiv a(\eta)a(\eta')H^2\Bigl[\Vert\vec{x}-\vec{x}'\Vert^2
-\left(\vert\eta-\eta'\vert-i\delta\right)^2\Bigr] \; . \label{LengthFuc}
\end{equation}

\subsection{Primitive diagrams}\label{PDs}

\begin{figure}
\hspace{1.9cm}
\vspace{-1.5cm}
\includegraphics[width=3.7cm,height=3cm]{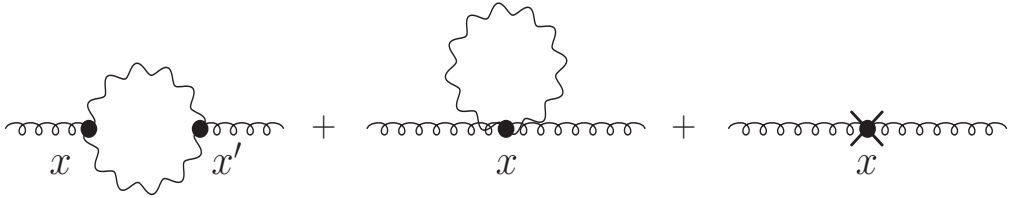}
\caption{Photon contributions to the one-loop graviton self-energy. Photon propagators are wavy and graviton propagators are curly.}
\label{gravgamma}
\end{figure}

All possible propagators and interaction vertices are derived from the primitive Lagrangian density $\mathcal{L}$,
\begin{equation}
\mathcal{L}=\mathcal{L}_{\text{GR}}+\mathcal{L}_{\text{EM}}+\mathcal{L}_{\text{GRfix}}+\mathcal{L}_{\text{EMfix}}+\mathcal{L}_{\text{BPHZ}} \; . \label{pmLag}
\end{equation}
Here, the gravitational Lagrangian density of low energy effective field theory is
\begin{equation}
\mathcal{L}_{\text{GR}}=\frac{1}{\kappa^2}\Bigl[R-(D-2)\Lambda\Bigr]\sqrt{-g} \; , \label{GRLag}
\end{equation}
where the factor of $(D-2)$ multiplying $\Lambda$ makes the pure gravitational field equation take the form of $R_{\mu\nu}=\Lambda g_{\mu\nu}$ in any dimension. And we define the cosmological constant as $\Lambda\equiv (D-1)H^2$. The Lagrangian density for electromagnetism takes the familiar form of Maxwell's theory,
\begin{equation}
\mathcal{L}_{\text{EM}}=-\frac{1}{4}F_{\mu\nu}F_{\rho\sigma}g^{\mu\rho}g^{\nu\sigma}\sqrt{-g} \; , \label{EMLag}
\end{equation}

\paragraph{\textbullet\; Propagators}\label{Props}
\ \\
\ \\
To get the graviton propagator, we expand the Lagrangian density (\ref{GRLag}) with the full metric (\ref{FullMetric}) and extract a presumably irrelevant surface term. Then, the quadratic part takes the form \cite{Woodard:2004ut}
\begin{eqnarray}
\lefteqn{\mathcal{L}_{\rm GR} \!-\! {\rm Surface\;Terms} = \left(\frac{D}{2}-1\right)Ha^{D-1} \sqrt{-\widetilde{g}}\,\widetilde{g}^{\rho\sigma} \widetilde{g}^{\mu\nu} h_{\nu 0}\partial_{\mu}h_{\rho\sigma} } \nonumber \\
& & \hspace{0cm} + a^{D-2} \sqrt{-\widetilde{g}}\,\widetilde{g}^{\alpha\beta}\widetilde{g}^{\rho\sigma}\widetilde{g}^{\mu\nu} \Biggl\{ \frac12 \partial_{\mu}h_{\alpha\rho} \partial_{\beta}h_{\nu\sigma} \!-\! \frac12 \partial_{\rho}h_{\alpha\beta}
\partial_{\nu}h_{\sigma\mu} \!+\! \frac14 \partial_{\rho}h_{\alpha\beta}
\partial_{\sigma}h_{\mu\nu} \!-\! \frac14 \partial_{\mu}h_{\alpha\rho}
\partial_{\nu}h_{\beta\sigma} \Biggr\} . \qquad \label{quaGRLag}
\end{eqnarray}
We fix the gauge by adding \cite{Woodard:2004ut}
\begin{equation}
\mathcal{L}_{\rm GRfix} = -\frac12 a^{D-2} \eta^{\mu\nu} F_{\mu}F_{\nu} \;\; , \;\; F_{\mu} \equiv \eta^{\rho\sigma} \left(\!\partial_{\sigma}h_{\mu\rho} - \frac12 \partial_{\mu}h_{\rho\sigma} + (D-2)Ha\delta_{\sigma}^0h_{\mu\rho}\!\right) \; . \label{GRfix}
\end{equation}
This gauge fixing term (\ref{GRfix}) is nice because it allow us to express the graviton propagator as a sum of scalar propagators multiplied by constant tensor factors,
\begin{equation}
i\bigl[{}_{\mu\nu}\Delta_{\alpha\beta}\bigr](x;x') \equiv \sum_{I=A,B,C} \Bigl[{}_{\mu\nu}T^I_{\alpha\beta}\Bigr] \times i\Delta_I(x;x') \; . \label{GravPropSum}
\end{equation}
The various tensor factors are
\begin{eqnarray}
\Bigl[{}_{\mu\nu} T^A_{\rho\sigma}\Bigr] & = & 2 \, \overline{\eta}_{\mu (\rho}
\overline{\eta}_{\sigma) \nu} - \frac2{(D\!-\! 3)} \overline{\eta}_{\mu\nu}
\overline{\eta}_{\rho \sigma} \; , \label{TA} \\
\Bigl[{}_{\mu\nu} T^B_{\rho\sigma}\Bigr] & = & -4 \delta^0_{(\mu}
\overline{\eta}_{\nu) (\rho} \delta^0_{\sigma)} \; , \label{TB} \\
\Bigl[{}_{\mu\nu} T^C_{\rho\sigma}\Bigr] & = & \frac2{(D \!-\!2) (D \!-\!3)}
\Bigl[(D \!-\!3) \delta^0_{\mu} \delta^0_{\nu} + \overline{\eta}_{\mu\nu}\Bigr]
\Bigl[(D \!-\!3) \delta^0_{\rho} \delta^0_{\sigma} + \overline{\eta}_{\rho
\sigma}\Bigr] \; , \label{TC}
\end{eqnarray}
where $\overline{\eta}^{\mu\nu} \equiv \eta^{\mu\nu} + \delta^{\mu}_0
\delta^{\nu}_0$ is the spatial part of the Minkowski metric. Each of those scalar propagators obey the associated propagator equation,
\begin{equation}
\mathcal{D}_I i\Delta_I(x;x') = i\delta^D(x-x') \qquad , \qquad I=A,B,C \; ,\label{ScPropEqn}
\end{equation}
where we define three scalar kinetic operators as follows:
\begin{equation}
\mathcal{D}_A \equiv \partial_{\mu}\left( \sqrt{-g}g^{\mu\nu}\partial_{\nu} \right) \equiv \sqrt{-g}\; \square\; , \label{ScDA}
\end{equation}
and
\begin{equation}
\mathcal{D}_B \equiv  \mathcal{D}_A - (D-2)H^2\sqrt{-g} \qquad , \qquad \mathcal{D}_C \equiv  \mathcal{D}_A - 2(D-3)H^2\sqrt{-g} \; . \label{ScDADB}
\end{equation}
Note that $i\Delta_A(x;x')$ is just the propagator of a MMC scalar. Thanks to previous work \cite{Tsamis:1992xa}, all of the scalar propagators have been worked out,
\begin{eqnarray}
\lefteqn{
i\Delta_A(x;x') = i\Delta_{\rm cf}(x;x')
} \nonumber\\
& & \hspace{0cm} 
+\frac{H^{D-2}}{(4\pi)^{\frac{D}{2}}} \frac{\Gamma(D-1)}{\Gamma(\frac{D}{2})} \Biggl\{ \frac{D}{D-4} \frac{\Gamma^2(\frac{D}{2})}{\Gamma(D-1)} \Bigl(\frac{4}{y}\Bigr)^{\frac{D}{2}-2} - \pi\cot\Bigl(\frac{\pi}{2} D\Bigr) + \ln(aa')\Biggr\} \nonumber\\
& & \hspace{0cm} 
+ \frac{H^{D-2}}{(4\pi)^{\frac{D}{2}}} \sum_{n=1}^{\infty} \Biggl\{ \frac1{n} \frac{\Gamma(n\!+\!D\!-\!1)}{\Gamma(n\!+\!\frac{D}{2})}\Bigl(\frac{y}{4}\Bigr)^n - \frac{1}{n\!-\!\frac{D}{2}\!+\!2} \frac{\Gamma(n\!+\!\frac{D}{2}\!+\!1)}{\Gamma(n\!+\!2)} \Bigl(\frac{y}{4}
\Bigr)^{n - \frac{D}2 +2} \Biggr\} \; , \label{DeltaA}\\
\lefteqn{
i \Delta_B(x;x') = B(y) = i\Delta_{\rm cf}(x;x')
} \nonumber\\
& & \hspace{0cm} 
- \frac{H^{D-2}}{(4\pi)^{\frac{D}{2}}} \sum_{n=0}^{\infty} \left\{ \frac{\Gamma(n+D-2)}{\Gamma(n+\frac{D}{2})} \Bigl(\frac{y}{4}\Bigr)^n - \frac{\Gamma(n+\frac{D}{2})}{\Gamma(n+2)} \Bigl( \frac{y}{4}\Bigr)^{n-\frac{D}{2}+2} \right\} \; , \label{DeltaB}\\
\lefteqn{
i\Delta_C(x;x') = C(y)= i\Delta_{\rm cf}(x;x')
} \nonumber\\
& & \hspace{0cm} 
+ \frac{H^{D-2}}{(4\pi)^{\frac{D}{2}}} \sum_{n=0}^{\infty} \Biggl\{ (n+1) \frac{\Gamma(n+D-3)}{\Gamma(n+\frac{D}{2})}\Bigl(\frac{y}{4}\Bigr)^n \nonumber\\
& & \hspace{5cm} 
- \Bigl(n-\frac{D}{2}+3\Bigr) \frac{\Gamma(n+\frac{D}{2}-1)}{\Gamma(n+2)} \Bigl(\frac{y}{4}\Bigr)^{n -\frac{D}{2}+2} \Biggr\} \; , \label{DeltaC}
\end{eqnarray}
where $i \Delta_{\rm cf}(x;x')$ is the de Sitter invariant propagator
of a conformally coupled scalar,
\begin{equation}
{i\Delta}_{\rm cf}(x;x') = \frac{H^{D-2}}{(4\pi)^{\frac{D}2}} \Gamma\left(
\frac{D}2 - 1\right) \left(\frac4{y}\right)^{\frac{D}2-1} \; .
\end{equation}
Note that the infinite series in all of these scalar propagators vanish in $D = 4$ dimensions, which means they are only needed when multiplied by some divergent pieces. 

Unlike gravitons, photons show no physical breaking of de Sitter invariance. So, we could in principle employ an exact de Sitter invariant gauge to make the photon propagator manifestly de Sitter invariant \cite{Tsamis:2006gj}. However, a noncovariant gauge fixing $\mathcal{L}_{\text{EMfix}}$ \cite{Woodard:2004ut,Kahya:2005kj} results in a much simpler form of the photon propagator,
\begin{equation}
i \bigl[{}_{\mu} \Delta_{\nu}\bigr](x;x') =
\overline{\eta}_{\mu\nu} \times a a' i\Delta_B(x;x') - \delta^0_{\mu}
\delta^0_{\nu} \times a a' i\Delta_C(x;x') \; . \label{Photprop}
\end{equation}
Here and henceforth we adopt a shorthand notation for the conformal time dependence of the scale factor,
\begin{equation}
a \equiv a(\eta) \;\; , \;\; a' \equiv a (\eta') \; .
\end{equation}
\paragraph{\textbullet\; Vertices}\label{Verts}
\ \\
\ \\
Note that the action of electromagnetism, $S_{\text{EM}}=\int\!\!d^Dx\mathcal{L}_{\text{EM}}$, contains interactions between two photons and any number of gravitons. So we obtain the three-point and four-point interaction vertices by functionally differentiating $S_{\text{EM}}$ once and twice with respect to the graviton field, and then setting the graviton field to zero,
\begin{eqnarray}
\frac{\delta S_{\text{EM}}}{\delta h_{\mu\nu}(x)} \biggl\vert_{h = 0} &=& \kappa a^{D-4}V^{\alpha\kappa\beta\lambda\mu\nu}\partial_{\kappa}A_{\alpha}(x)\partial_{\lambda}A_{\beta}(x) \; , \label{3PtVert} \\
\frac{\delta^2 S_{\text{EM}}}{\delta h_{\mu\nu}(x)\delta h_{\rho\sigma}(x')} \biggl\vert_{h = 0} &=&\kappa^2 a^{D-4}U^{\alpha\kappa\beta\lambda\mu\nu\rho\sigma}\partial_{\kappa}A_{\alpha}(x)\partial_{\lambda}A_{\beta}(x)\delta^D(x-x') \; . \label{4PtVert}
\end{eqnarray}
The tensor factors for the three-point and four-point vertices are \cite{Leonard:2013xsa}
\begin{eqnarray}
\lefteqn{V^{\alpha\kappa\beta\lambda\mu\nu}
=\eta^{\mu\nu}\eta^{\alpha[\beta}\eta^{\lambda]\kappa}
+4\eta^{\mu)[\alpha}\eta^{\kappa][\beta}\eta^{\lambda](\nu} \; ,} \label{Vvert} \\
\lefteqn{U^{\alpha\kappa\beta\lambda\mu\nu\rho\sigma}
=\Bigl(\frac14\eta^{\mu\nu}\eta^{\rho\sigma}-\frac12\eta^{\mu(\rho}\eta^{\sigma)\nu}\Bigr)\eta^{\alpha[\beta}\eta^{\lambda]\kappa}
} \nonumber \\
& & \hspace{0cm} 
+\eta^{\mu\nu}\eta^{\rho)[\alpha}\eta^{\kappa][\beta}\eta^{\lambda](\sigma}
+\eta^{\rho\sigma}\eta^{\mu)[\alpha}\eta^{\kappa][\beta}\eta^{\lambda](\nu} \nonumber \\
& & \hspace{0cm}
+\eta^{\alpha(\mu}\eta^{\nu)(\rho}\eta^{\sigma)[\beta}\eta^{\lambda]\kappa}
+\eta^{\kappa(\mu}\eta^{\nu)(\rho}\eta^{\sigma)[\lambda}\eta^{\beta]\alpha}
\nonumber \\
& & \hspace{0cm}
+\eta^{\alpha(\mu}\eta^{\nu)[\beta}\eta^{\lambda](\rho}\eta^{\sigma)\kappa}
+\eta^{\kappa(\mu}\eta^{\nu)[\lambda}\eta^{\beta](\rho}\eta^{\sigma)\alpha} \nonumber \\
& & \hspace{0cm}
+\eta^{\alpha[\beta}\eta^{\lambda](\mu}\eta^{\nu)(\rho}\eta^{\sigma)\kappa}
+\eta^{\kappa[\lambda}\eta^{\beta](\mu}\eta^{\nu)(\rho}\eta^{\sigma)\alpha} \; .  \label{Uvert}
\end{eqnarray}

\paragraph{\textbullet\; Diagram Expressions}\label{DiagExpr}
\ \\
\ \\
Now, using the Feynman rules stated in section \ref{FRs}, we can write down the formal expressions for the first two diagrams in Fig. \ref{gravgamma}. The leftmost diagram, also known as the ``three-point function'', is constructed from two three-point vertices,
\begin{eqnarray}
\lefteqn{-i\bigl[{}^{\mu\nu}\Sigma^{\rho\sigma}_{\text{3pt}}\bigr](x;x')
=\frac12
\left(-i\kappa\right)a^{D-4}V^{\alpha\kappa\gamma\theta\mu\nu}\partial_{\kappa}\partial'_{\lambda}i\bigl[{}_{\alpha}\Delta_{\beta}\bigr](x;x')} \nonumber\\
& & \hspace{4cm} \times\left(-i\kappa\right)a'^{D-4}V^{\beta\lambda\delta\phi\rho\sigma}\partial_{\theta}\partial'_{\phi}i\bigl[{}_{\gamma}\Delta_{\delta}\bigr](x;x') \; . \label{3ptExpr}
\end{eqnarray}
The middle diagram, also known as the ``four-point function'', is constructed from a single four-point vertex,
\begin{equation}
-i\bigl[{}^{\mu\nu}\Sigma^{\rho\sigma}_{\text{4pt}}\bigr](x;x')
=\left(-i\kappa^2\right)a^{D-4}U^{\alpha\kappa\beta\lambda\mu\nu\rho\sigma}\partial_{\kappa}\partial'_{\lambda}i\bigl[{}_{\alpha}\Delta_{\beta}\bigr](x;x')\delta^D(x-x') \; . \label{4ptExpr}
\end{equation}
The rightmost diagram denotes a series of counterterms which will be explained in section \ref{CtTerms}.

\subsection{Structure functions}\label{dSStrcFun}
The graviton self-energy possesses 100 components in $D=4$ dimensions. We could report on all of them, but the symmetries will relate these components such that they can be expressed using a few structure functions. This idea is actually inspired by experience in flat space. For example, the Poincar\'e invariance makes the vacuum polarization of photons take the form of a single structure function $\Pi\bigl((x-x')^2\bigr)$,
\begin{equation}
i\bigl[{}^{\mu}\Pi^{\nu}_{\text{flat}}\bigr](x;x')
=\left(\eta^{\mu\nu}\eta^{\rho\sigma}-\eta^{\mu\sigma}\eta^{\nu\rho}\right)\partial_{\rho}\partial'_{\sigma}i\Pi\bigl((x-x')^2\bigr) \; .
\end{equation}
And the graviton self-energy takes the form
\begin{equation}
-i\bigl[{}^{\mu\nu}\Sigma^{\rho\sigma}_{\text{flat}}\Bigr](x;x')
=\Pi^{\mu\nu}\Pi^{\rho\sigma}F_0\bigl((x-x')^2\bigr)
\!+\!\left(\Pi^{\mu(\rho}\Pi^{\sigma)\nu}-\frac{\Pi^{\mu\nu}\Pi^{\rho\sigma}}{D-1}\right)F_2\bigl((x-x')^2\bigr) \; , \label{flatF+G}
\end{equation}
where we define the projection operator as $\Pi^{\mu\nu}\equiv \partial^{\mu}\partial^{\nu}-\eta^{\mu\nu}\partial^2$. We usually call $F_0$ and $F_2$ as the spin zero and spin two part. The de Sitter background for inflation is spatially homogeneous and isotropic. Then, if these are the only isometries, we take it to be manifest in our representation for the graviton self-energy that it turns out that we shall need four structure functions. And guided by the previous studies \cite{Leonard:2012si,Leonard:2012ex} about a noncovariant representation of the vacuum polarization on de Sitter spacetime,
\begin{equation}
i\bigl[{}^{\mu}\Pi^{\nu}\bigr](x;x')
=\left(\eta^{\mu\nu}\eta^{\rho\sigma}\!-\!\eta^{\mu\sigma}\eta^{\nu\rho}\right)\partial_{\rho}\partial'_{\sigma}F(x;x')
+\left(\overline{\eta}^{\mu\nu}\overline{\eta}^{\rho\sigma}\!-\!\overline{\eta}^{\mu\sigma}\overline{\eta}^{\nu\rho}\right)\partial_{\rho}\partial'_{\sigma}G(x;x')
\; ,
\end{equation}
which makes interpretation of the physics much easier. Here, we will adopt a similar noncovariant representation of the graviton self-energy \cite{Leonard:2014zua},
\begin{eqnarray}
\lefteqn{-i\bigl[{}^{\mu\nu}\Sigma^{\rho\sigma}\bigr](x;x')
=\mathcal{F}^{\mu\nu}(x)\times\mathcal{F}^{\rho\sigma}(x')\Bigl[F_0(x;x')\Bigr]} \nonumber\\
& & \hspace{0cm}
+\mathcal{G}^{\mu\nu}(x)\times\mathcal{G}^{\rho\sigma}(x')\Bigl[G_0(x;x')\Bigr]
+\mathcal{F}^{\mu\nu\rho\sigma}\Bigl[F_2(x;x')\Bigr]
+\mathcal{G}^{\mu\nu\rho\sigma}\Bigl[G_2(x;x')\Bigr] \; . \label{F+G}
\end{eqnarray}
where two scalar projectors, $\mathcal{F}^{\mu\nu}$ and $\mathcal{G}^{\mu\nu}$, are
\begin{eqnarray}
\lefteqn{
\mathcal{F}^{\mu\nu}=\partial^{\mu}\partial^{\nu}
+2(D-1)aH\delta^{(\mu}_0\partial^{\nu)} +(D-2)(D-1)a^2H^2\delta^{\mu}_0\delta^{\nu}_0
} \nonumber\\
& & \hspace{4cm}
-\eta^{\mu\nu}\left[\partial^2+(D-1)aH\partial_0+(D-1)a^2H^2\right] \; , \label{SFProj}\\
\lefteqn{
\mathcal{G}^{\mu\nu}=\overline{\partial}^{\mu}\overline{\partial}^{\nu}
+2(D-2)aH\delta^{(\mu}_0\overline{\partial}^{\nu)} +(D-2)(D-1)a^2H^2\delta^{\mu}_0\delta^{\nu}_0
} \nonumber\\
& & \hspace{4cm}
-\overline{\eta}^{\mu\nu}\left[\nabla^2+(D-2)aH\partial_0+(D-2)a^2H^2\right] \; , \label{SGProj}
\end{eqnarray}
and where $\overline{\partial}^{\mu}\equiv\partial^{\mu}+\delta^{\mu}_0\partial_0$. And two tensor projectors, $\mathcal{F}^{\mu\nu\rho\sigma}$ and $\mathcal{G}^{\mu\nu\rho\sigma}$, are
\begin{eqnarray}
\mathcal{F}^{\mu\nu\rho\sigma} &=& \mathcal{C}_{\alpha\beta\gamma\delta}{}^{\mu\nu}(x)
\times\mathcal{C}_{\kappa\lambda\theta\phi}{}^{\rho\sigma}(x')
\times\eta^{\alpha\kappa}\eta^{\beta\lambda}\eta^{\gamma\theta}\eta^{\delta\phi} \; , \label{TFProj}\\
\mathcal{G}^{\mu\nu\rho\sigma} &=& \mathcal{C}_{\alpha\beta\gamma\delta}{}^{\mu\nu}(x)
\times\mathcal{C}_{\kappa\lambda\theta\phi}{}^{\rho\sigma}(x')
\times\overline{\eta}^{\alpha\kappa}\overline{\eta}^{\beta\lambda}\overline{\eta}^{\gamma\theta}\overline{\eta}^{\delta\phi} \; , \label{TGProj}
\end{eqnarray}
where we define the second order differential operator $\mathcal{C}_{\alpha\beta\gamma\delta}{}^{\mu\nu}$ by expanding the Weyl tensor of the conformally transformed metric to the linear order of the graviton field,
\begin{equation}
\widetilde{g}_{\mu\nu}\equiv\eta_{\mu\nu}+\kappa h_{\mu\nu} \;\;\;\; \Rightarrow \;\;\;\; \widetilde{C}_{\alpha\beta\gamma\delta}\equiv\mathcal{C}_{\alpha\beta\gamma\delta}{}^{\mu\nu}\times\kappa h_{\mu\nu}+O(\kappa^2h^2) \; . \label{WeylOptr}
\end{equation}
Explicit expressions for $\mathcal{F}^{\mu\nu\rho\sigma}$ and $\mathcal{G}^{\mu\nu\rho\sigma}$ can be found in the Appendix of \cite{Leonard:2014zua}.

It is worth mentioning that the de Sitter representation (\ref{F+G}) must be consistent with its corresponding flat space limit (\ref{flatF+G}). This means that all terms in $G_0(x;x')$ and $G_2(x;x')$ must contain at least one factor of $H^2$, so as to vanish in the limit that $H$ vanishes with the comoving time $t = \ln(a)/H$ held fixed. Hence, the leading divergences are contained in $F_0(x;x')$ and $F_2(x;x')$. $G_0(x;x')$ and $G_2(x;x')$ are correspondingly less divergent. 

Now this leads to the question, how do we find the structure functions? In general, suppose that after substituting the photon propagators in our formal expressions given by Eqs. (\ref{3ptExpr}) and (\ref{4ptExpr}), doing index contractions and acting derivatives, we obtain a primitive result for $-i\bigl[{}^{\mu\nu}\Sigma^{\rho\sigma}\bigr](x;x')$. Since this result can be expressed in the form suggested by Eq. (\ref{F+G}), as we explained previously, we can extract out those four structure functions by looking at particular components. The procedure will be as follows.
\begin{itemize}
\item{First, we take the trace on one index group of $-i\bigl[{}^{\mu\nu}\Sigma^{\rho\sigma}\bigr]$, say $\rho$ and $\sigma$. This will cause the spin two part to drop out. Then we shall look at the components $-i\bigl[{}^{0i}\Sigma^{\rho\sigma}\bigr]\times\eta_{\rho\sigma}$ and $-i\bigl[{}^{jk}\Sigma^{\rho\sigma}\bigr]\times\eta_{\rho\sigma}$ ($j\neq k$) to obtain two linearly independent equations to solve for $F_0(x;x')$ and $G_0(x;x')$.}
\item{Next, we look at the components $-i\bigl[{}^{0i}\Sigma^{jk}\bigr]$ and $-i\bigl[{}^{jk}\Sigma^{0i}\bigr]$, and always assume that $i\neq j\neq k\neq i$. This will also allow us to derive two linearly independent equations to solve for $F_2(x;x')$ and $G_2(x;x')$.}
\end{itemize}
More details about how this procedure works will be given when we carry out the explicit calculation in section \ref{Spin0StrcFun} and \ref{Spin2StrcFun}.

\section{ONE-LOOP GRAVITON SELF-ENERGY}\label{GravSE}
\subsection{Primitive results of three-point and four-point functions}\label{PrimRes}
In this section, we will substitute our vertex tensor factors (\ref{Vvert}) and (\ref{Uvert}) into the formal expressions (\ref{3ptExpr}) and (\ref{4ptExpr}) and then perform naive contractions. Before diving into the tensor algebra, it is worth noticing the fact that in both of the diagram expressions (\ref{3ptExpr}-\ref{4ptExpr}) the doubly differentiated photon propagators inherit an antisymmetry from the vertex tensor factors through index contractions. So, we will have a cleaner, simpler analysis if expressing the computation in terms of a four-index object which is the correlator of two field strengths,
\begin{equation}
i \bigl[{}_{\alpha\beta} \Delta_{\gamma\delta}\bigr](x;x') = \langle\Omega\vert F_{\alpha\beta}(x)F_{\gamma\delta}(x') \vert\Omega\rangle = 4 \partial_{[\alpha}\partial'_{[[\gamma} i \bigl[{}_{\beta]} \Delta_{\delta]]}\bigr](x;x') \; . \label{FFcor}
\end{equation}
Here, the single and double square brackets indicate antisymmetrization on the index pairs $\alpha\leftrightarrow\beta$ and $\gamma\leftrightarrow\delta$, respectively.

Now let us evaluate the field strength correlator (\ref{FFcor}). Starting with our photon propagator expression (\ref{Photprop}), we want to first express the tensor factors in terms of the derivatives of $y(x;x')$ and $u(x;x') \equiv \ln(aa')$,
\begin{eqnarray}
aa'\overline{\eta}_{\mu\nu} & = & 
\frac1{2H^2} \Bigl[ -\partial_{\mu}\partial'_{\nu}y 
+ \partial_{\mu}y\times\partial'_{\nu}u 
+ \partial_{\mu}u \times \partial'_{\nu}y 
+ (2-y)\partial_{\mu}u \times \partial'_{\nu}u \Bigr] \; , \label{aaetabar} \\
aa'\delta_{\mu}^{0}\delta_{\nu}^{0} & = & 
\frac1{2H^2} \Bigl[ 2\partial_{\mu}u \times \partial'_{\nu}u \Bigr] \; . \label{aadelta}
\end{eqnarray}
And there is a very useful relation between $B(y)$ and $C(y)$,
\begin{equation}
C(y) = \frac12 (2-y) B(y) + \frac{k}{(D-3)} \qquad \text{where} \qquad k \equiv \frac{H^{D-2}}{(4\pi)^{\frac{D}2}} \frac{\Gamma(D-1)}{\Gamma(\frac{D}2)} \; . \label{CtoB}
\end{equation}
Substituting expressions (\ref{aaetabar})-(\ref{CtoB}) into (\ref{Photprop}) allows one to express our propagator as
\begin{equation}
i \bigl[{}_{\mu} \Delta_{\nu}\bigr](x;x') =
-\frac1{2H^2} \Bigl( \partial_{\mu}\partial'_{\nu}y 
- \partial_{\mu}y \partial'_{\nu}u 
- \partial_{\mu}u \partial'_{\nu}y \Bigr) B(y)
- \frac{k}{(D-3)H^2} \Bigl( \partial_{\mu}u \partial'_{\nu}u \Bigr) \; . \label{Photprop2}
\end{equation}
The next step is to act two derivatives on (\ref{Photprop2}) and antisymmetrize them to form the field strength correlator (\ref{FFcor}). Notice the fact that $\partial_{\mu}u=Ha\delta_{\mu}^0$ is independent of $x'^{\mu}$ and that $\partial'_{\nu}u=Ha'\delta_{\nu}^0$ is independent of $x^{\mu}$. It follows that only the de Sitter invariant first term of expression (\ref{Photprop2}) contributes to the field strength correlator,
\begin{equation}
i \bigl[{}_{\alpha\beta} \Delta_{\gamma\delta}\bigr](x;x') = -\frac2{H^2} \Bigl( D_{[\alpha}D'_{[[\gamma}B(y) \Bigr) \Bigl( D_{\beta]}D'_{\delta]]}y \Bigr) \; , \label{invFFcor}
\end{equation}
where $D_{\alpha}$ and $D'_{\gamma}$ are the covariant derivatives associated with the de Sitter background metrics $\overline{g}_{\mu\nu}(x) \equiv a^2\eta_{\mu\nu}$ and $\overline{g}_{\rho\sigma}(x') \equiv a'^2\eta_{\rho\sigma}$, respectively. It also turns out to be useful to work out the coincidence limit of the field strength correlator. Starting with (\ref{invFFcor}) and using the fact that
\begin{equation}
\lim_{x \to x'} \Bigl( D_{\alpha}D'_{\gamma}B(y) \Bigr) = -2H^2 a'^2 B'(0) \eta_{\alpha\gamma} \qquad \text{where} \qquad B'(0)\equiv \lim_{y \to 0} \Bigl( \frac{d}{dy}B(y) \Bigr)\; . \label{coinddB}
\end{equation}
Hence, the coincident field strength correlator is
\begin{equation}
\lim_{x \to x'} i \bigl[{}_{\alpha\beta} \Delta_{\gamma\delta}\bigr](x;x')  = -8H^2 a'^4 B'(0) \eta_{\alpha[\gamma}\eta_{\delta]\beta} \; . \label{coinFFcor}
\end{equation}
The final identity we would like to derive for future convenience is based on the fact there are, of course, local delta function contributions hidden in the expression (\ref{invFFcor}) because of the doubly differentiated scalar propagator. And one can show that
\begin{equation}
D_{\mu}D'_{\nu}B(y) = \Bigl\{ D_{\mu}D'_{\nu}B(y) \Bigr\}_{\text{naive}} + \frac1{a^{D-2}} \delta_{\mu}^0 \delta_{\nu}^0 i\delta^D(x-x') \; , \label{ddBdelta}
\end{equation}
where $\left\{\ldots\right\}_{\text{naive}}$ denotes whatever we got by naively acting derivatives. Then, the result (\ref{ddBdelta}) allows us to segregate out the local delta function contributions in the field strength correlator,
\begin{equation}
i \bigl[{}_{\alpha\beta} \Delta_{\gamma\delta}\bigr](x;x')  = \Bigl\{ i \bigl[{}_{\alpha\beta} \Delta_{\gamma\delta}\bigr](x;x') \Bigr\}_{\text{naive}} + \frac4{a^{D-4}}\delta_{[\beta}^0 \eta_{\alpha][\gamma} \delta_{\delta]}^0 i\delta^D(x-x') \; . \label{FFcordelta}
\end{equation}
Now we are all set to compute the primitive results of the first two diagrams in Fig. \ref{gravgamma} by substitution of (\ref{Vvert}) and (\ref{Uvert}) in (\ref{3ptExpr}) and (\ref{4ptExpr}) and naive index contractions. First, it is easy to see that in terms of the field strength correlator (\ref{FFcor}) the four-point and three-point functions can be rewritten as
\begin{eqnarray}
\lefteqn{
-i\bigl[{}^{\mu\nu}\Sigma^{\rho\sigma}_{\text{4pt}}\bigr](x;x')
= -\frac12 \kappa^2 a'^{D-4} i\delta^D(x-x') \times
} \nonumber\\
& & \hspace{-0.5cm}
\Biggl\{\!\! \left(\!\frac18 \eta^{\mu\nu}\eta^{\rho\sigma} 
\!-\! \frac14 \eta^{\mu(\rho}\eta^{\sigma)\nu} \!\right)
i\bigl[{}^{\alpha\beta} \Delta_{\alpha\beta}\bigr](x';x')
\!-\! \frac12 \eta^{\mu\nu} i\bigl[{}^{\rho\beta} \Delta^{\sigma}{}_{\beta}\bigr](x';x')
\!-\! \frac12 \eta^{\rho\sigma} i\bigl[{}^{\mu\beta} \Delta^{\nu}{}_{\beta}\bigr](x';x') \nonumber\\
& & \hspace{5.3cm}
+ 2 \eta^{\nu)(\rho} i\bigl[{}^{\sigma)\beta} \Delta^{(\mu}{}_{\beta}\bigr](x';x')
- i\bigl[{}^{\mu(\rho} \Delta^{\sigma)\nu}\bigr](x';x')
\Biggr\} \; , \label{4ptFF} \\
\lefteqn{
-i\bigl[{}^{\mu\nu}\Sigma^{\rho\sigma}_{\text{3pt}}\bigr](x;x')
= -\frac12 \kappa^2 (aa')^{D-4} \times
} \nonumber\\
& & \hspace{-0.5cm}
\Biggl\{\! \frac1{16} \eta^{\mu\nu}\eta^{\rho\sigma} i\bigl[{}^{\alpha\beta} \Delta^{\gamma\delta}\bigr](x;x')\!  \times \! i\bigl[{}_{\alpha\beta} \Delta_{\gamma\delta}\bigr](x;x') 
\! -\!  \frac14 \eta^{\mu\nu} i\bigl[{}^{\alpha\beta} \Delta^{\gamma\rho}\bigr](x;x')\!  \times \! i\bigl[{}_{\alpha\beta} \Delta_{\gamma}{}^{\sigma}\bigr](x;x') \nonumber\\
& & \hspace{-0.3cm}
- \frac14 \eta^{\rho\sigma} i\bigl[{}^{\alpha\mu} \Delta^{\gamma\delta}\bigr](x;x')\!  \times \! i\bigl[{}_{\alpha}{}^{\nu} \Delta_{\gamma\delta}\bigr](x;x')
\! +\! i\bigl[{}^{\alpha\mu} \Delta^{\gamma(\rho}\bigr](x;x')\!  \times \! i\bigl[{}_{\alpha}{}^{\nu} \Delta_{\gamma}{}^{\sigma)}\bigr](x;x') \Biggl\} \; . \label{3ptFF}
\end{eqnarray}
Note that while getting (\ref{4ptFF}) and (\ref{3ptFF}) we raise or lower indices only with the Minkowski metric, which just plays with the relative minus sign between the temporal and spatial components.

\paragraph{\textbullet\; Local contributions}
\ \\
\ \\
Note that all terms in (\ref{4ptFF}) are proportional to the delta function. So, applying the coincident identity of the field strength correlator (\ref{coinFFcor}), we immediately obtain the contributions from our four-point functions which are purely local,
\begin{eqnarray}
\lefteqn{
-i\bigl[{}^{\mu\nu}\Sigma^{\rho\sigma}_{\text{4pt}}\bigr]_{\text{local}}(x;x') = 
-\kappa^2 H^2 a'^D B'(0) \Biggl\{ 
\frac12 \left( D^2-9D+12 \right) \eta^{\mu(\rho}\eta^{\sigma)\nu} 
} \nonumber\\
& & \hspace{5cm}
- \frac14 \left( D^2-9D+16 \right) \eta^{\mu\nu}\eta^{\rho\sigma}
\Biggr\} i\delta^D(x-x') \; . \label{4ptlocal}
\end{eqnarray}
Obviously, the expression (\ref{3ptFF}) will contain both local and nonlocal contributions, and we first deal with the local terms. So, applying the identity (\ref{FFcordelta}) and ignoring all of the naive terms for the time being, we can also easily obtain the local contributions from our three-point function,
\begin{eqnarray}
\lefteqn{
-i\bigl[{}^{\mu\nu}\Sigma^{\rho\sigma}_{\text{3pt}}\bigr]_{\text{local}}(x;x') = 
-\kappa^2 H^2 a'^D B'(0) \Biggl\{ 
\left( D-5 \right) \eta^{\mu\nu}\eta^{\rho\sigma} 
+ 4 \eta^{\mu(\rho}\eta^{\sigma)\nu}
} \nonumber\\
& & \hspace{-0.5cm}
+ 2\! \left( D-4 \right) \eta^{\mu\nu}\delta_0^{\sigma}\delta_0^{\rho}
\!+\! 2\! \left( D-4 \right) \eta^{\rho\sigma}\delta_0^{\mu}\delta_0^{\nu}
\!-\! 4\! \left( D-4 \right) \delta_0^{(\mu}\eta^{\nu)(\rho}\delta_0^{\sigma)} \Biggr\} i\delta^D(x-x') \; . \label{3ptlocal}
\end{eqnarray}
Adding (\ref{4ptlocal}) and (\ref{3ptlocal}) together simply gives the entire local contributions from our three-point and four-point functions,
\begin{eqnarray}
\lefteqn{
-i\bigl[{}^{\mu\nu}\Sigma^{\rho\sigma}\bigr]_{\text{local}}(x;x') = 
(D-4)\kappa^2 H^2 a'^D B'(0) \Biggl\{ 
\frac14 (D-9) \eta^{\mu\nu}\eta^{\rho\sigma} 
} \nonumber\\
& & \hspace{0cm}
- \frac12 (D-5) \eta^{\mu(\rho}\eta^{\sigma)\nu}
- 2 \eta^{\mu\nu}\delta_0^{\sigma}\delta_0^{\rho}
- 2 \eta^{\rho\sigma}\delta_0^{\mu}\delta_0^{\nu}
+ 4 \delta_0^{(\mu}\eta^{\nu)(\rho}\delta_0^{\sigma)} \Biggr\} i\delta^D(x-x') \; . \label{4pt3ptlocal}
\end{eqnarray}
From our explicit expression of $B(y)$ (\ref{DeltaB}), we know that $B'(0)$ is finite in $D = 4$ dimensions,
\begin{equation}
B'(0) = -\frac{H^{D-2}}{4(4\pi)^{\frac{D}2}} \frac{\Gamma(D-1)}{\Gamma(\frac{D}2+1)} \; . \label{dB0}
\end{equation}
So, we can simply set $D = 4$ and conclude that the
net contribution of (\ref{4pt3ptlocal}) vanishes.

\paragraph{\textbullet\; Nonlocal contributions}
\ \\
\ \\
Now let us focus on the nonlocal contributions which are nothing but all of the naive terms, that produce no delta functions, in our 3pt function. Please note that (\ref{3ptFF}) is not written in a manifestly covariant form because we have raised indices using the Minkowski inverse metric instead of the inverse background metric. That is merely for the purpose of naive index contraction, but we can easily recover the covariance by putting back the appropriate factors of $a$ or $a'$. Also, note that the tensor structure of the naive terms in our field strength correlator (\ref{invFFcor}) can be expressed in terms of the covariant basis furnished by derivatives of $y(x;x')$,
\begin{eqnarray}
\lefteqn{
\Bigl\{ i \bigl[{}_{\alpha\beta} \Delta_{\gamma\delta}\bigr](x;x') \Bigr\}_{\text{naive}} = \frac2{H^2} \Bigl[ \left( D_{[\alpha}y \right) \left( D_{\beta]}D'_{[\gamma}y \right) \left( D'_{\delta]}y \right) B''(y)
} \nonumber\\
& & \hspace{7cm}
- \left( D_{\alpha}D'_{[\gamma}y \right) \left( D'_{\delta]}D_{\beta}y \right) B'(y) \Bigr] \; . \label{naiveFFcor}
\end{eqnarray} 
And one can show there are some very useful contraction identities of these basis tensors \cite{Kahya:2005kj},
\begin{eqnarray}
\overline{g}^{\mu\nu}(x) \left(D_{\mu}y\right) \left(D_{\nu}y\right) & = & H^2 \bigl(4 y - y^2\bigr) =
\overline{g}^{\rho\sigma}(x') \left(D'_{\rho}y\right) \left(D'_{\sigma}y\right) \; ,
\label{contraction1}\\
\overline{g}^{\mu\nu}(x) \left(D_{\mu}y\right) \left(D_{\nu}D'_{\rho}y\right) & = & H^2 (2-y) \left(D'_{\rho}y\right) \; ,
\label{contraction2}\\
\overline{g}^{\rho\sigma}(x') \left(D'_{\rho}y\right) \left(D_{\mu}D'_{\sigma}y\right) & = & H^2 (2-y) \left(D_{\mu}y\right) \; ,
\label{contraction3}\\
\overline{g}^{\mu\nu}(x) \left(D_{\mu}D'_{\rho}y\right) \left(D_{\nu}D'_{\sigma}y\right) & = & 4 H^4 \overline{g}_{\rho\sigma}(x') - H^2 \left(D'_{\rho}y\right) \left(D'_{\sigma}y\right) \; ,
\label{contraction4}\\
\overline{g}^{\rho\sigma}(x') \left(D_{\mu}D'_{\rho}y\right) \left(D_{\nu}D'_{\sigma}y\right) & = & 4 H^4 \overline{g}_{\mu\nu}(x) - H^2 \left(D_{\mu}y\right) \left(D_{\nu}y\right) \; .
\label{contraction5}
\end{eqnarray}
We remind the reader that $\overline{g}^{\mu\nu}=a^{-2}\eta^{\mu\nu}$ and $\overline{g}^{\rho\sigma}=a'^{-2}\eta^{\rho\sigma}$ are the inverse background metrics on the de Sitter background. Next step is to substitute our result (\ref{naiveFFcor}) for the naive terms in the expression (\ref{3ptFF}), and then to make use of the contraction identities (\ref{contraction1})-(\ref{contraction5}). It is straightforward to express all of the naive terms in our three-point function as a linear combination of the de Sitter invariant basis tensors,
\begin{eqnarray}
\lefteqn{
-i\bigl[{}^{\mu\nu}\Sigma^{\rho\sigma}\bigr](x;x') = 
(aa')^{D+2} \biggl\{ \left( D^{\mu}D'^{(\rho}y \right) \left( D'^{\sigma)}D^{\nu}y \right) \alpha(y)
} \nonumber\\
& & \hspace{-0.5cm}
\!+\! \left( D^{(\mu}y \right)\!\! \left( D^{\nu)}D'^{(\rho}y \right)\!\! \left( D'^{\sigma)}y \right)\! \beta(y)
\!+\! \left( D^{\mu}y \right)\!\! \left( D^{\nu}y \right)\!\! \left( D'^{\rho}y \right)\!\! \left( D'^{\sigma}y \right)\! \gamma(y) 
+ H^4 \overline{g}^{\mu\nu}(x) \overline{g}^{\rho\sigma}(x') \delta(y) \nonumber\\
& & \hspace{2.5cm}
+ H^2 \Bigl[ \overline{g}^{\mu\nu}(x) \left( D'^{(\rho}y \right) \left( D'^{\sigma)}y \right) + \overline{g}^{\rho\sigma}(x') \left( D^{(\mu}y \right) \left( D^{\nu)}y \right) \Bigr] \epsilon (y) \biggr\} \; . \label{3ptnonlocal}
\end{eqnarray}
And those five scalar functions in (\ref{3ptnonlocal}) are given by,
\begin{eqnarray}
\alpha(y) \equiv \lefteqn{
\kappa^2 \Biggl\{ -\frac18 \left(4y-y^2\right)^2 \bigl[B''(y)\bigr]^2
-\biggl[2(D-2)-\frac12\left(4y-y^2\right)\biggr]\bigl[B'(y)\bigr]^2
} \nonumber\\
& & \hspace{4.8cm}
-\frac12(2-y)\left(4y-y^2\right)\bigl[B'(y)B''(y)\bigr]
\Biggl\} \; ,
\label{scalara}\\
\beta(y) \equiv \lefteqn{
\kappa^2 \Biggl\{ \frac14 (2-y)\left(4y-y^2\right) \bigl[B''(y)\bigr]^2
-(2-y)\bigl[B'(y)\bigr]^2
} \nonumber\\
& & \hspace{4cm}
-\biggl[2(D-4)+\left(4y-y^2\right)\biggr]\bigl[B'(y)B''(y)\bigr]
\Biggl\} \; ,
\label{scalarb}\\
\gamma(y) \equiv \lefteqn{
\kappa^2 \Biggl\{ -\biggl[\frac12(D-2)-\frac18\left(4y-y^2\right)\biggr]\bigl[B''(y)\bigr]^2
-\frac12 \bigl[B'(y)\bigr]^2
} \nonumber\\
& & \hspace{6.7cm}
+\frac12 (2-y) \bigl[B'(y)B''(y)\bigr]
\Biggl\}
\; ,
\label{scalarc}
\end{eqnarray}
\begin{eqnarray}
\delta(y) \equiv \lefteqn{
\kappa^2 \Biggl\{\!\! -\frac18(D-5)\!\!\left(4y-y^2\right)\!\!^2\bigl[B''(y)\bigr]^2\!
\!-\!\biggl[\!\left(D^2\!-\!9D\!+\!16\right)\!\!-\!\frac12(D\!-\!5)\!\!\left(4y\!-\!y^2\right)\!\biggr]\!\bigl[B'(y)\bigr]^2
} \nonumber\\
& & \hspace{3.5cm}
-\frac12(D-5)(2-y)\left(4y-y^2\right)\bigl[B'(y)B''(y)\bigr]
\Biggl\}
\; ,
\label{scalard}\\
\epsilon(y) \equiv \lefteqn{
\kappa^2 \Biggl\{ 
\frac14 (D-4)\left(4y-y^2\right) \bigl[B''(y)\bigr]^2
-(D-4)\bigl[B'(y)\bigr]^2
} \nonumber\\
& & \hspace{5.7cm}
+(D-4)(2-y)\bigl[B'(y)B''(y)\bigr]
\Biggl\} \; .
\label{scalare}
\end{eqnarray}
Since our local contributions (\ref{4pt3ptlocal}) vanishes at one-loop order, the nonlocal terms (\ref{3ptnonlocal}) complete the one-loop photon contributions to graviton self-energy.

\subsection{Spin zero structure functions}\label{Spin0StrcFun}
In this section, we will follow the procedure introduced in section \ref{dSStrcFun} to extract two spin zero structure functions $F_0(x;x')$ and $G_0(x;x')$. We start with the trace of expression (\ref{3ptnonlocal}),
\begin{eqnarray}
\lefteqn{
-i\bigl[{}^{\mu\nu}\Sigma^{\rho\sigma}\bigr](x;x') \times \eta_{\rho\sigma} = a^2H^2(aa')^D \Biggl\{
H^2 \overline{g}^{\mu\nu}(x) 
\Bigl[4\alpha(y)+D\delta(y)+\left(4y-y^2\right)\epsilon(y)\Bigr]
} \nonumber\\
& & \hspace{0.5cm}
+ \left( D^{\mu}y \right) \left( D^{\nu}y \right) 
\Bigl[-\alpha(y)+(2-y)\beta(y)+\left(4y-y^2\right)\gamma(y)+D\epsilon(y)\Bigr] \Biggr\} \; . \label{tr3ptnonlocal}
\end{eqnarray}
Now, recall that our strategy is to look at the components $-i\bigl[{}^{0i}\Sigma^{\rho\sigma}\bigr]\times\eta_{\rho\sigma}$ and $-i\bigl[{}^{jk}\Sigma^{\rho\sigma}\bigr]\times\eta_{\rho\sigma}$ ($j\neq k$) to infer equations that $F_0$ and $G_0$ satisfy. One can see that the time derivative on $y(x;x')$ gives
\begin{equation}
\partial_0 y = aH\left(y-2+2\frac{a'}{a}\right) \; , \label{d0y}
\end{equation}
which leads to the components we need of $\left( D^{\mu}y \right) \left( D^{\nu}y \right)$,
\begin{eqnarray}
\left( D^0 y \right) \left( D^iy \right) & = & 
-a^{-4}\left(\partial_0 y\right)\left(\partial_iy\right) 
= -a^{-4}\times aH\left(y-2+2\frac{a'}{a}\right)\left(\partial_iy\right) \; , \label{d0ydiy} \\
\left( D^j y \right) \left( D^ky \right) & = & 
a^{-4} \times \left( \partial_jy \right)\left( \partial_ky \right) \; . \label{djydky}
\end{eqnarray}
As explained in work \cite{Leonard:2014zua}, now we need to recognize two important scalar sources $S_1(x;x')$ and $S_2(x;x')$, by writing
\begin{eqnarray}
-i\bigl[{}^{0i}\Sigma^{\rho\sigma}\bigr](x;x')\times\eta_{\rho\sigma} & \equiv & 
a^{D-2} a'^{D} \times \partial_i S_1(x;x') \; , \label{0iSigma} \\
-i\bigl[{}^{jk}\Sigma^{\rho\sigma}\bigr](x;x')\times\eta_{\rho\sigma} & \equiv & 
a^{D-2} a'^{D} \times \partial_j \partial_k S_2(x;x') \; . \label{jkSigma}
\end{eqnarray}
If we define the indefinite integral of a function $f(y)$ as $I[f(y)] \equiv \int^ydzf(z),$ and another scalar function $F(y)$ as
\begin{equation}
F(y) \equiv I^2 \Bigl[ -\alpha(y)+(2-y)\beta(y)+\left(4y-y^2\right)\gamma(y)+D\epsilon(y) \Bigr]
 \; , \label{scalarF}
\end{equation}
one can see that
\begin{eqnarray}
-i\bigl[{}^{0i}\Sigma^{\rho\sigma}\bigr](x;x')\times\eta_{\rho\sigma} & = & 
a^{D-2} a'^{D} \partial_i \Bigl\{\! 
-aH^3 I\Bigl[\!(y-2)F''(y)\!\Bigr] \!-\! 2a'H^3 F'(y) \!\Bigr\} \; , \label{0iSigma2} \\
-i\bigl[{}^{jk}\Sigma^{\rho\sigma}\bigr](x;x')\times\eta_{\rho\sigma} & = & 
a^{D-2} a'^{D} \partial_j \partial_k \Bigl\{H^2 F(y)\Bigr\}  \; . \label{jkSigma2}
\end{eqnarray}
It follows that
\begin{eqnarray}
S_1(x;x') & = & H^2(-\partial_0+aH)F(y) \; , \label{S1} \\
S_2(x;x') & = & H^2 F(y) \; . \label{S2}
\end{eqnarray}
Then, substituting (\ref{S1})  and (\ref{S2}) into (56) and (57) of \cite{Leonard:2014zua} allows us to express the spin zero structure functions as
\begin{eqnarray}
F_0(x;x') & = & -\frac{(aa')^{D-2}}{(D-1)} \left(\frac{H^2}{\square'+DH^2}\right) F(y) \; , \label{F0eq} \\
G_0(x;x') & = & 0 \; , \label{G0eq}
\end{eqnarray}
where $\square'$ denotes the invariant scalar d'Alembertian evaluated at $x'^{\mu}$, i.e., $\square' \equiv \overline{g}^{\rho\sigma}(x')D'_{\rho}D'_{\sigma} = a'^{-D} \partial_{\rho}\left(a'^{D-2}\eta^{\rho\sigma}\partial_{\sigma}\right)$. Now, let us first work out what explicitly the scalar source $F(y)$ is. Substituting (\ref{scalara})-(\ref{scalarc}) and (\ref{scalare}) into (\ref{scalarF}) gives
\begin{equation}
F(y) = \kappa^2 (D-2)(D-4) I^2 \Bigl[ -\left(B'\right)^2 
+ (2-y)\left(B'B''\right) + \frac14\left(4y-y^2\right)\left(B''\right)^2 \Bigr]
 \; . \label{scalarFexpr}
\end{equation}
Then, substituting the explicit expression of $B(y)$ (\ref{DeltaB}) leads to the following expansion of $F(y)$:
\begin{equation}
F(y) = \frac{(D-4)^2}{16(D-1)} \frac{\kappa^2 H^{2D-4} \Gamma^2(\frac{D}2)}{(4\pi^D)} \Biggl\{
(D-2)\left(\frac4{y}\right)^{D-1}
- (D-4)\left(\frac4{y}\right)^{D-2} + \ldots \Biggr\}
 \; , \label{scalarFser}
\end{equation}
where we have neglected terms that are integrable and vanish in $D=4$ dimensions. Similar terms will also be regarded as irrelevant in subsequent expansions. 

The next step is, of course, to recover the structure function $F_0(x;x')$ from expression (\ref{F0eq}) by employing the Green's function for $\left(\square'+DH^2\right)$. But note that electromagnetism is conformally invariant in $D=4$ dimensions, which means that our graviton self-energy shall be traceless in those dimensions; therefore, the spin zero structure function $F_0(x;x')$ is expected to contain a factor of $(D-4)$, as we see in our result of source term $F(y)$ (\ref{scalarFser}). However, there happens to be a factor of $(D-4)^2$ in $F(y)$ which causes absolute zero one-loop contributions to $F_0(x;x')$ since the most divergence we can have at this order is $(D-4)^{-1}$. To sum up, both of the spin zero structure functions, $F_0(x;x')$ and $G_0(x;x')$, vanish at one-loop order.

\subsection{Spin two structure functions}\label{Spin2StrcFun}
In this section, we intend to extract the spin two structure functions $F_2(x;x')$ and $G_2(x;x')$ from the components $-i\bigl[{}^{0i}\Sigma^{jk}\bigr](x;x')$ and $-i\bigl[{}^{jk}\Sigma^{0i}\bigr](x;x')$, with $i\neq j\neq k\neq i$ \cite{Leonard:2014zua},
\begin{eqnarray}
-i\bigl[{}^{0i}\Sigma^{jk}\bigr](x;x') & \equiv & 
(aa')^{D-2}\partial_i \partial_j \partial_k S_3(x;x') \; , \label{0iSigmajk} \\
-i\bigl[{}^{jk}\Sigma^{0i}\bigr](x;x') & \equiv & 
(aa')^{D-2}\partial_i \partial_j \partial_k S_4(x;x')  \; . \label{jkSigma0i}
\end{eqnarray}
This means that we will need the following identities:
\begin{eqnarray}
\left(D^0 y\right)\left(D^i y\right) & = & -a^{-4} \times aH \left( y-2+2\frac{a'}{a} \right)  \left(\partial_i y\right) \; , \label{S2id1} \\
\left(D'^0 y\right)\left(D'^i y\right) & = & a'^{-4} \times a'H \left( y-2+2\frac{a}{a'} \right)\left(\partial_i y\right) \; , \label{S2id2}
\end{eqnarray}
\begin{equation}
\left(D^0 D'^i y\right) = (aa')^{-2} \times aH \left(\partial_i y\right) 
\;\;\; , \;\;\; 
\left(D'^0 D^i y\right) = -(aa')^{-2} \times a'H \left(\partial_i y\right) \; . \label{S2id3}
\end{equation}
Applying these identities to the desired components of (\ref{3ptnonlocal}) gives
\begin{eqnarray}
-i\bigl[{}^{0i}\Sigma^{jk}\bigr](x;x') & = & 
(aa')^{D-2}\partial_i \partial_j \partial_k \biggl\{\!
-\frac12 HaI^3\Bigl[\beta(y)\Bigr] \!\!-\! \left(\partial_0\!-\!3Ha\right)I^4\Bigl[\gamma(y)\Bigr] \!\biggr\}
 \; , \label{0iSigmajk2} \\
-i\bigl[{}^{jk}\Sigma^{0i}\bigr](x;x') & = & 
(aa')^{D-2}\partial_i \partial_j \partial_k \biggl\{\!
\frac12 Ha'I^3\Bigl[\beta(y)\Bigr] \!\!+\! \left(\partial'_0\!-\!3Ha'\right)I^4\Bigl[\gamma(y)\Bigr]\! \biggr\}  \; . \label{jkSigma0i2}
\end{eqnarray}
Then, comparing the results (\ref{0iSigmajk2}) and (\ref{jkSigma0i2}) with expressions (\ref{0iSigmajk}) and (\ref{jkSigma0i}) gives another two scalar sources,
\begin{eqnarray}
S_3(x;x') & = & -\frac12 HaI^3\Bigl[\beta(y)\Bigr] - \left(\partial_0-3Ha\right)I^4\Bigl[\gamma(y)\Bigr] \; , \label{S3} \\
S_4(x;x') & = & \frac12 Ha'I^3\Bigl[\beta(y)\Bigr] + \left(\partial'_0-3Ha'\right)I^4\Bigl[\gamma(y)\Bigr] \; . \label{S4}
\end{eqnarray}
Substituting expressions (\ref{S3}) and (\ref{S4}) into (66) and (67) of \cite{Leonard:2014zua} and taking into account that in our particular case both $F_0(x;x')$ and $G_0(x;x')$ vanish, we find that $F_2(x;x')$ and $G_2(x;x')$ are determined by the following two independent equations:
\begin{eqnarray}
\lefteqn{
\left(\partial'_0-\partial_0\right)\Biggl\{
\frac{(D-3)}{(D-1)}F_2
-\frac{(D-3)}{(D-1)(D-2)^2}G_2-(aa')^{D-2}I^4\Bigl[\gamma(y)\Bigr] \Biggr\}
} \nonumber\\
& & \hspace{1cm}
= (a-a')H(aa')^{D-2}\times\biggl\{
-\frac12 I^3\Bigl[\beta(y)\Bigr]+(D+1)I^4\Bigl[\gamma(y)\Bigr]\biggr\} \; , \label{F2G2eqn1} \\
\lefteqn{
\left(\partial'_0+\partial_0\right)\Biggl\{
\frac{(D-3)D}{(D-1)(D-2)}F_2
+\frac{(D-3)}{(D-1)(D-2)^2}G_2+(aa')^{D-2}I^4\Bigl[\gamma(y)\Bigr] \Biggr\}
} \nonumber\\
& & \hspace{1cm}
= (a+a')H(aa')^{D-2}\times\biggl\{
-\frac12 I^3\Bigl[\beta(y)\Bigr]+(D+1)I^4\Bigl[\gamma(y)\Bigr]\biggr\} \; . \label{F2G2eqn2}
\end{eqnarray}
To solve equation (\ref{F2G2eqn1}), consider acting $(\partial'_0-\partial_0)$ on $(aa')^{D-2}f(y)$,
\begin{equation}
(\partial'_0-\partial_0)(aa')^{D-2}f(y)
= (a-a')H(aa')^{D-2}\Bigl[(4-y)f'(y)-(D-2)f(y)\Bigr] \; .
\end{equation} 
If we let
\begin{equation}
(aa')^{D-2}f(y) \equiv \frac{(D-3)}{(D-1)}F_2
-\frac{(D-3)}{(D-1)(D-2)^2}G_2-(aa')^{D-2}I^4\Bigl[\gamma(y)\Bigr] \; ,
\end{equation} 
then (\ref{F2G2eqn1}) becomes a first order linear differential equation,
\begin{equation}
f'(y)-\frac{(D-2)}{(4-y)}f(y) = \frac1{(4-y)}\biggl\{
-\frac12 I^3\Bigl[\beta(y)\Bigr]+(D+1)I^4\Bigl[\gamma(y)\Bigr]\biggr\} \; ,
\end{equation} 
which allows us to invert $(\partial'_0-\partial_0)$ in (\ref{F2G2eqn1}) and obtain an indefinite integral,
\begin{eqnarray}
\lefteqn{
\frac{(D-3)}{(D-1)}F_2(x;x') -\frac{(D-3)}{(D-1)(D-2)^2}G_2(x;x')
-(aa')^{D-2}I^4\Bigl[\gamma(y)\Bigr] 
} \nonumber\\
& & \hspace{-0.5cm}
= \frac{(aa')^{D-2}}{(4-y)^{D-2}} \Biggl\{
I\biggl[(4-y)^{D-3}\biggl\{-\frac12 I^3\Bigl[\beta(y)\Bigr]+(D+1)I^4\Bigl[\gamma(y)\Bigr]\biggr\}\biggr]+K_1\Biggr\} \; , \label{F2G2eqn1sol}
\end{eqnarray}
where the integration constant $K_1$ is chosen so that there is no singularity at the antipodal point $y=4$. Similarly, to solve Eq. (\ref{F2G2eqn2}), consider acting $(\partial'_0+\partial_0)$ on $(aa')^{D-2}f(y)$,
\begin{equation}
(\partial'_0+\partial_0)(aa')^{D-2}f(y)
= (a+a')H(aa')^{D-2}\Bigl[yf'(y)+(D-2)f(y)\Bigr] \; ,
\end{equation} 
which allows us to invert $(\partial'_0+\partial_0)$ in (\ref{F2G2eqn2}) and obtain another indefinite integral,
\begin{eqnarray}
\lefteqn{
\frac{(D-3)D}{(D-1)(D-2)}F_2(x;x') +\frac{(D-3)}{(D-1)(D-2)^2}G_2(x;x')
+(aa')^{D-2}I^4\Bigl[\gamma(y)\Bigr] 
} \nonumber\\
& & \hspace{1cm}
= \frac{(aa')^{D-2}}{y^{D-2}} \Biggl\{
I\biggl[y^{D-3}\biggl\{-\frac12 I^3\Bigl[\beta(y)\Bigr]+(D+1)I^4\Bigl[\gamma(y)\Bigr]\biggr\}\biggr]+K_2\Biggr\} \; , \label{F2G2eqn2sol}
\end{eqnarray}
where the integration constant $K_2$ is chosen so as to keep $G_2(x;x')$ less divergent than $F_2(x;x')$. Equations (\ref{F2G2eqn1sol}) and (\ref{F2G2eqn2sol}) simply imply that
\begin{eqnarray}
F_2(x;x') &=& \frac{(D-2)}{2(D-3)}\Biggl\{
\frac{(aa')^{D-2}}{(4-y)^{D-2}}P(y)+\frac{(aa')^{D-2}}{y^{D-2}}Q(y)\Biggr\} \; , \label{F2sol1}\\
G_2(x;x') &=& \frac{(D-1)(D-2)^2}{(D-3)}
\lefteqn{ \Biggl\{\frac{(D-3)}{(D-1)}F_2(x;x')} \nonumber\\
& & \hspace{3cm}
-(aa')^{D-2}I^4\Bigl[\gamma(y)\Bigr]-\frac{(aa')^{D-2}}{(4-y)^{D-2}}P(y)\Biggr\} \; . \label{G2sol1}
\end{eqnarray}
where we have defined two functions of $y(x;x')$,
\begin{eqnarray}
P(y) & \equiv & I\biggl[(4-y)^{D-3}\biggl\{-\frac12 I^3\Bigl[\beta(y)\Bigr]+(D+1)I^4\Bigl[\gamma(y)\Bigr]\biggr\}\biggr]+K_1 \; , \label{Py}\\
Q(y) & \equiv & I\biggl[y^{D-3}\biggl\{-\frac12 I^3\Bigl[\beta(y)\Bigr]+(D+1)I^4\Bigl[\gamma(y)\Bigr]\biggr\}\biggr]+K_2 \; . \label{Qy}
\end{eqnarray}
Substituting the explicit expression of $B(y)$ (\ref{DeltaB}) into (\ref{scalarb}) and (\ref{scalarc}) and working out various indefinite integrals, we obtain the following expansions:
\begin{eqnarray}
I^3\Bigl[\beta(y)\Bigr] & = & \frac{\kappa^2 H^{2D-4} \Gamma^2(\frac{D}2)}{(4\pi)^D} \Biggl\{-\frac{(3D-8)}{2(D-2)(D-1)}\left(\frac4{y}\right)^{D-2}+\ldots\Biggr\}\; , \label{I3beta}\\
I^4\Bigl[\gamma(y)\Bigr] & = & \frac{\kappa^2 H^{2D-4} \Gamma^2(\frac{D}2)}{(4\pi)^D} \Biggl\{-\frac{D}{8(D-1)(D+1)}\left(\frac4{y}\right)^{D-2}+\ldots\Biggr\} \; . \label{I4gamma}
\end{eqnarray}
As always, the neglected terms in these expansions are integrable and vanish in $D = 4$ dimensions. Thanks to the factor of $(D-4)^2$ in the following result, which is a little surprising,
\begin{eqnarray}
\lefteqn{
-\frac12 I^3\Bigl[\beta(y)\Bigr]+(D+1)I^4\Bigl[\gamma(y)\Bigr]
} \nonumber \\
& & \hspace{1.5cm}
= \frac{\kappa^2 H^{2D-4} \Gamma^2(\frac{D}2)}{(4\pi)^D} \Biggl\{-\frac{(D-4)^2}{8(D-2)(D-1)}\left(\frac4{y}\right)^{D-2}+\ldots\Biggr\}\; , \label{I3bI4g}
\end{eqnarray}
our analysis becomes a lot easier. As a result, we find
\begin{equation}
P(y) = K_1 \;\;\; , \;\;\; Q(y) = K_2 \; . \label{PyQysol}
\end{equation}
Note that we shall set $K_1=0$ to remove the singularity at the antipodal point $y=4$. Then, it follows that
\begin{eqnarray}
F_2(x;x') &=& \frac{(D-2)}{2(D-3)}\Biggl\{
\frac{(aa')^{D-2}}{y^{D-2}}K_2\Biggr\} \; , \label{F2sol2}\\
G_2(x;x') &=& \frac{(D-1)(D-2)^2}{(D-3)}\lefteqn{ \Biggl\{\frac{(D-2)}{2(D-1)}
\frac{(aa')^{D-2}}{y^{D-2}}K_2} \nonumber\\
& & \hspace{0.5cm}
+ \frac{\kappa^2 (H^2aa')^{D-2} \Gamma^2(\frac{D}2)}{(4\pi)^D} \frac{D}{8(D-1)(D+1)}\left(\frac4{y}\right)^{D-2} \Biggr\}\; . \label{G2sol2}
\end{eqnarray}
Note that (\ref{F2sol2}) and (\ref{G2sol2}) suggest that both $F_2$ and $G_2$ are proportional to $1/y^{D-2}$. Recall that $G_2$ should be less singular than $F_2$: therefore, the integration constant $K_2$ can only be chosen so that $G_2$ vanishes at this order,
\begin{equation}
\frac{(aa')^{D-2}}{y^{D-2}}K_2 =  \frac{\kappa^2 (H^2aa')^{D-2} \Gamma^2(\frac{D}2)}{(4\pi)^D} \Biggl\{
-\frac{D}{4(D-2)(D+1)}\left(\frac4{y}\right)^{D-2} \Biggr\}\; . \label{K2}
\end{equation}
It follows that
\begin{eqnarray}
F_2(x;x') & = & \frac{\kappa^2 (H^2aa')^{D-2} \Gamma^2(\frac{D}2)}{(4\pi)^D} \Biggl\{
-\frac{D}{8(D-3)(D+1)}\left(\frac4{y}\right)^{D-2} \Biggr\}\; , \label{F2sol3}\\
G_2(x;x') & = & 0 \; . \label{G2sol3}
\end{eqnarray}

\section{RENORMALIZATION}\label{Renorm}
\subsection{Counterterms}\label{CtTerms}
To figure out what that appropriate BPHZ counterterms are, we are guided by two facts, the first of which is that we are calculating a 1PI graviton two-point function. This means our counterterms should contain two graviton fields. The second fact is the superficial degree of divergence of 4 in our primitive diagrams at one-loop order, which means that the counterterms should contain up to four derivatives. For quantum gravity at one-loop order, the necessary counterterms can be taken to be the Ricci curvature square $R^2$ and Weyl tensor square $C^{\alpha\beta\gamma\delta}C_{\alpha\beta\gamma\delta}$ \cite{'tHooft:1974bx}. Note that we have obtained a perfectly de Sitter invariant representation of our graviton self-energy (\ref{3ptnonlocal}), thanks to the one-loop photon contribution being de Sitter invariant. As a result, there would be no additional counterterms.

Moreover, it has been shown in section (\ref{Spin0StrcFun}-\ref{Spin2StrcFun}) that the nontrivial contributions of photons to the graviton self-energy come from the spin two part. Thus, the only counterterm we really need is the Weyl tensor square counterterm which is defined by the following Lagrangian:
\begin{equation}
\Delta\mathcal{L}_2 \equiv c_2 C^{\alpha\beta\gamma\delta}C_{\alpha\beta\gamma\delta} \sqrt{-g}
\; . \label{DelL2}
\end{equation}
Using the fact that the Weyl tensor with one upper index is conformally invariant,
\begin{equation}
C^{\alpha}{}_{\beta\gamma\delta} = \widetilde{C}^{\alpha}{}_{\beta\gamma\delta}\;\;\;\; \Rightarrow \;\;\;\; C_{\alpha\beta\gamma\delta} = a^2\widetilde{C}_{\alpha\beta\gamma\delta} \; . \label{Weylid}
\end{equation}
We can rewrite the counterterm Lagrangian (\ref{DelL2}) as
\begin{eqnarray}
\Delta\mathcal{L}_2 &=& c_2 a^{D-4} \widetilde{C}^{\alpha\beta\gamma\delta}\widetilde{C}_{\alpha\beta\gamma\delta} \sqrt{-\widetilde{g}} \nonumber\\
&=& c_2 a^{D-4} \eta^{\alpha\kappa}\eta^{\beta\lambda}\eta^{\gamma\theta}\eta^{\delta\phi}
\left(\mathcal{C}_{\alpha\beta\gamma\delta}{}^{\mu\nu}\kappa h_{\mu\nu}\right)
\left(\mathcal{C}_{\kappa\lambda\theta\phi}{}^{\rho\sigma}\kappa h_{\rho\sigma}\right) + O\left(\kappa^3h^3\right)
\; . \label{DelL2con}
\end{eqnarray}
Note that we have used the linearized Weyl tensor operator $\mathcal{C}_{\alpha\beta\gamma\delta}{}^{\mu\nu}$ defined in section \ref{dSStrcFun} to expand the counterterm Lagrangian up to quadratic order in the graviton field. Then, one obtains the counterterm vertex by functionally differentiating $i$ times counterterm action twice, and then setting the graviton field to zero,
\begin{equation}
\frac{i\delta^2 \Delta S_2}{\delta h_{\mu\nu}(x)\delta h_{\rho\sigma}(x')}\Biggl\vert_{h = 0} = 2c_2 \kappa^2 \eta^{\alpha\kappa}\eta^{\beta\lambda}\eta^{\gamma\theta}\eta^{\delta\phi}
\mathcal{C}_{\alpha\beta\gamma\delta}{}^{\mu\nu}
{\mathcal{C}'}_{\kappa\lambda\theta\phi}{}^{\rho\sigma} 
\!\times\! \Bigl[ a^{D-4} i\delta^D(x-x') \Bigr] \; . \label{CTMvert}
\end{equation}
By recalling the definition (\ref{TFProj}) of $\mathcal{F}^{\mu\nu\rho\sigma}$ one can see 
that the contribution of the Weyl tensor square counterterm to our graviton self-energy is
\begin{equation}
-i\bigl[{}^{\mu\nu}\Delta\Sigma^{\rho\sigma}_2\bigr](x;x') = 2c_2\kappa^2 \mathcal{F}^{\mu\nu\rho\sigma} \!\times\! \Bigl[a^{D-4} i\delta^D(x-x')\Bigr] 
\; . \label{CTMSE}
\end{equation}

\subsection{Renormalizing the nonintegrability}\label{renormF2}
In the previous section, we derived our graviton self-energy in arbitrary $D$ dimensions,
\begin{equation}
-i\bigl[{}^{\mu\nu}\Sigma^{\rho\sigma}\bigr](x;x') = \mathcal{F}^{\mu\nu\rho\sigma} \Bigl[F_2(x;x')\Bigr] \; , \label{SEprimexpr}
\end{equation}
where $F_2(x;x')$ is given by (\ref{F2sol3}). One sees that our primitive result (\ref{SEprimexpr}) contains the factor of $y^{2-D}$ which diverges at coincidence ($x^{\mu}=x'^{\mu}$) with nonzero coefficients in $D=4$ dimensions. Recall that to quantum correct the linearized Einstein field equation (\ref{lineffeq}) we will integrate $-i\bigl[{}^{\mu\nu}\Sigma^{\rho\sigma}\bigr]$ over the four-dimensional measure; hence, (\ref{SEprimexpr}) is not integrable in those dimensions. However, only the terms that are at least as singular as $y^{-2}$ need to be renormalized. Less singular terms are actually integrable and therefore can be evaluated in $D = 4$. To renormalize those nonintegrable terms, the general procedure is
\begin{itemize}
\item{First, convert them into functions of $\Delta x^2$ using the definition $y\equiv aa'H^2\Delta x^2$.}

\item{Second, extract derivatives with respect to the unprimed coordinate $x^{\mu}$, for example,
\begin{eqnarray}
\frac{1}{\Delta x^{2D-4}} & = & \frac{\partial^2}{2(D-3)(D-4)}\frac{1}{\Delta x^{2D-6}} \; , \label{RenormId1} \\
\frac{1}{\Delta x^{2D-2}} & = & \frac{\partial^4}{4(D-2)^2(D-3)(D-4)}\frac{1}{\Delta x^{2D-6}} \; , \label{RenormId2}
\end{eqnarray}
in which the derivatives can be simply pulled out of the integration over primed coordinate $x'^{\mu}$ in the quantum corrected effective field equation (\ref{lineffeq}). We keep doing this so as to get a less and less singular integrand until it becomes integrable in $D=4$, but the price we pay for it is to get a manifest divergence, the factor of $(D-4)^{-1}$}.

\item{Third, localize the explicit divergence of $(D-4)^{-1}$} in the previous step by adding zero using the identity
\begin{equation}
\partial^2\frac{1}{\Delta x^{D-2}}=\frac{i4\pi^{D/2}}{\Gamma\left(\frac{D}{2}-1\right)}\delta^D(x-x') \; . \label{RenormId3}
\end{equation}
This means that we can write
\begin{eqnarray}
\frac{\partial^2}{(D-4)}\frac{1}{\Delta x^{2D-6}}
= \lefteqn{
\frac{i4\pi^{D/2}}{\Gamma\left(\frac{D}{2}-1\right)}\frac{\mu^{D-4}\delta^D(x-x')}{(D-4)} } \nonumber\\
& & \hspace{1cm}
+\frac{\partial^2}{(D-4)}\Biggl\{\frac{1}{\Delta x^{2D-6}}-\frac{\mu^{D-4}}{\Delta x^{D-2}}\Biggr\} \; , \label{RenormId4}
\end{eqnarray}
where the factor of $\mu$ with inverse length dimension is added for dimensional consistency. Then we expand the terms in the curly brackets away from $D=4$,
\begin{equation}
\frac{1}{\Delta x^{2D-6}}-\frac{\mu^{D-4}}{\Delta x^{D-2}}
=-\frac{(D-4)}{2}\frac{\ln\left(\mu^2\Delta x^2\right)}{\Delta x^2}+O\Bigl[(D-4)^2\Bigr] \; .
\end{equation}
So, we can write Eq. (\ref{RenormId4}) as
\begin{eqnarray}
\frac{\partial^2}{(D-4)}\frac{1}{\Delta x^{2D-6}}
=\lefteqn{
\frac{i4\pi^{D/2}}{\Gamma\left(\frac{D}{2}-1\right)}\frac{\mu^{D-4}\delta^D(x-x')}{(D-4)}}\nonumber\\
& & \hspace{1cm}
-\frac{\partial^2}{2}\Biggl[\frac{\ln\left(\mu^2\Delta x^2\right)}{\Delta x^2}\Biggr]+O(D-4) \; . \label{RenormId5}
\end{eqnarray}
Note that in Eq. (\ref{RenormId5}) the logarithm term is integrable and the delta function piece is local though singular in $D=4$. This is what we mean by localizing the divergence.

\item{Finally, use the technique introduced above to break our result of graviton self-energy into finite and divergent parts. Then we shall set appropriate counterterm coefficients so as to cancel all the divergent parts, and obtain the fully renormalized result.}
\end{itemize}
Now, we are all set to renormalize the primitive result of $-i\bigl[{}^{\mu\nu}\Sigma^{\rho\sigma}\bigr]$ (\ref{SEprimexpr}) actually, the primitive result of $F_2(x;x')$ (\ref{F2sol3}), to be exact. Applying the identities (\ref{RenormId1}) and (\ref{RenormId5}) in (\ref{F2sol3}) gives
\begin{equation}
F_2(x;x') = \frac{\kappa^2}{640\pi^4}\partial^2 \Biggl[\frac{\ln(\mu^2 \Delta x^2)}{\Delta x^2}\Biggr]-\frac{\kappa^2 \mu^{D-4}\Gamma(\frac{D}2)}{128\pi^{\frac{D}2}} \frac{D(D-2)i\delta^D(x-x')}{(D-4)(D-3)^2(D+1)} \; . \label{F2finidiv}
\end{equation}
Substituting (\ref{F2finidiv}) into (\ref{SEprimexpr}) and comparing it with (\ref{CTMSE}), one sees that the local divergent part will be completely subtracted off by the counterterm contribution if we choose the arbitrary counterterm coefficient $c_2$ to be
\begin{equation}
c_2 = \frac{\mu^{D-4}\Gamma(\frac{D}2)}{256\pi^{\frac{D}2}} \frac{D(D-2)}{(D-4)(D-3)^2(D+1)} \; . \label{c2}
\end{equation}
Now, using this counterterm coefficient $c_2$, we finally obtain the renormalized form of our graviton self-energy,
\begin{eqnarray}
-i\bigl[{}^{\mu\nu}\Sigma_{\text{re}}^{\rho\sigma}\bigr](x;x') & = & -i\bigl[{}^{\mu\nu}\Sigma^{\rho\sigma}\bigr](x;x') -i\bigl[{}^{\mu\nu}\Delta\Sigma_2^{\rho\sigma}\bigr](x;x') \nonumber \\
& = & \mathcal{F}^{\mu\nu\rho\sigma} \Bigl[ \frac{\kappa^2}{80\pi^2} \ln(a) i\delta^4(x-x') 
+ F_{2,\text{re}}(x;x')\Bigr]  \; . \label{SErenorm}
\end{eqnarray}
And the nonlocal part of our renormalized spin two structure function is
\begin{equation}
F_{2,\text{re}}(x;x') = \frac{\kappa^2}{640\pi^4}\partial^2 \Biggl[\frac{\ln(\mu^2 \Delta x^2)}{\Delta x^2}\Biggr] \; . \label{F2renorm}
\end{equation}
It is worth mentioning at this point that the spacetime dependence of equation (\ref{F2renorm}) allows us to easily reflect any derivatives on it, $\partial'\rightarrow-\partial$. This fact facilitates a dramatic simplification of the tensor projector $\mathcal{F}^{\mu\nu\rho\sigma}$ when acting on $F_{2,\text{re}}(\Delta x^2)$ in (\ref{SErenorm}), which will be be useful in section \ref{Source}. To be specific, given the explicit expression for the projector $\mathcal{F}^{\mu\nu\rho\sigma}$ in the Appendix of \cite{Leonard:2014zua}, one can see that in $D = 4$ dimensions it is reduced to the nice flat space form
\begin{equation}
\mathcal{F}^{\mu\nu\rho\sigma} \rightarrow \frac12 \left( \Pi^{\mu(\rho} \Pi^{\sigma)\nu} - \frac13 \Pi^{\mu\nu} \Pi^{\rho\sigma} \right) \;\;\; , \;\;\; \Pi^{\mu\nu} \equiv \partial^{\mu}\partial^{\nu} - \eta^{\mu\nu}\partial^2 \; . \label{Fproj}
\end{equation}
Before closing this section, please note another fact, that with the following two identities,
\begin{eqnarray}
\frac{1}{\Delta x^2} & = & \frac{\partial^2}{4} \ln\left(\mu^2 \Delta x^2\right) \; , \label{del2log1} \\
\frac{\ln\left(\mu^2 \Delta x^2\right)}{\Delta x^2}
& = & \frac{\partial^2}{8}\Bigl[ \ln^2\left(\mu^2 \Delta x^2\right) -2\ln\left(\mu^2 \Delta x^2\right) \Bigr]  \; , \label{del2log2}
\end{eqnarray}
we can achieve a significant simplification by extracting another d'Alembertian from (\ref{F2renorm})
\begin{equation}
F_{2,\text{re}}(x;x') = \frac{\kappa^2}{5120\pi^4} \partial^4 \Bigl[ \ln^2\left(\mu^2 \Delta x^2\right) -2\ln\left(\mu^2 \Delta x^2\right) \Bigr] \; . \label{F2renorm2}
\end{equation}
This form of the structure function (\ref{F2renorm2}) will come in handy when used in the Schwinger-Keldysh formalism.

\section{SOLVING THE EFFECTIVE FIELD EQUATION}\label{SolEffeqn}
\subsection{Schwinger-Keldysh formalism}\label{SKform}
The in-out result of the graviton self-energy (\ref{SErenorm}) seems to be ready for use in the linearized effective field equation (\ref{lineffeq}), but that would lead to two problems:
\begin{itemize}
\item{Causality -- The in-out graviton self-energy (\ref{SErenorm}) is nonzero for points $x'^{\mu}$ which lie in the future of $x^{\mu}$, or at spacelike separation from it; and}
\item{Reality -- The in-out graviton self-energy (\ref{SErenorm}) is not real.}
\end{itemize}
Note that there is nothing wrong with correcting the graviton propagator with the in-out graviton self-energy for asymptotic scattering amplitudes in flat space. But the causality and reality require that in the effective field equations, the appropriate 1PI two-point function should be that of the Schwinger-Keldysh (S-K) formalism. This technique provides a way of computing causal and real vacuum expectation values that is almost as simple as the Feynman diagrams which produce in-out matrix elements \cite{Schwinger:1960qe, Mahanthappa:1962ex, Bakshi:1962dv, Bakshi:1963bn, Keldysh:1964ud, Chou:1984es, Jordan:1986ug, Calzetta:1986ey}. 
In our particular case, the photon propagator depends upon the $y(x;x')$ which is a function of the Lorentz interval $\Delta x^2(x;x')$. The four Schwinger-Keldysh propagators can be obtained by making the following replacements for the Lorentz interval \cite{Ford:2004wc}:
\begin{eqnarray}
\Delta x^2_{++} \equiv \Bigl\Vert \vec{x} \!-\! \vec{x}' 
\Bigr\Vert^2 - \Bigl(\vert \eta \!-\! \eta'\vert \!-\! i \delta \Bigr)^2 & , &
\Delta x^2_{+-} \equiv \Bigl\Vert \vec{x} \!-\! \vec{x}' 
\Bigr\Vert^2 - \Bigl(\eta \!-\! \eta' \!+\! i \delta \Bigr)^2 , \nonumber \\
\Delta x^2_{--} \equiv \Bigl\Vert \vec{x} \!-\! \vec{x}' 
\Bigr\Vert^2 - \Bigl(\vert \eta \!-\! \eta'\vert \!+\! i \delta \Bigr)^2 & , &
\Delta x^2_{-+} \equiv \Bigl\Vert \vec{x} \!-\! \vec{x}' 
\Bigr\Vert^2 - \Bigl(\eta \!-\! \eta' \!-\! i \delta \Bigr)^2 . \qquad \label{SKDeltax}
\end{eqnarray}
To resolve the problem of reality and causality of the in-out formalism while used in an effective field equation, what we should really be using in (\ref{lineffeq}) is the retarded graviton self-energy $\bigl[{}^{\mu\nu}\Sigma_{\text{R}}^{\rho\sigma}\bigr]$ of the Schwinger-Keldysh formalism,
\begin{equation}
\bigl[{}^{\mu\nu}\Sigma^{\rho\sigma}\bigr](x;x') \rightarrow \bigl[{}^{\mu\nu}\Sigma_{\text{R}}^{\rho\sigma}\bigr](x;x') \equiv \bigl[{}^{\mu\nu}\Sigma_{++}^{\rho\sigma}\bigr](x;x')+\bigl[{}^{\mu\nu}\Sigma_{+-}^{\rho\sigma}\bigr](x;x') \; . \label{SKSE1}
\end{equation}
It's actually very simple to convert our in-out result (\ref{SErenorm}) into a S-K retarded one. We will get $\bigl[{}^{\mu\nu}\Sigma_{++}^{\rho\sigma}\bigr]$ and $\bigl[{}^{\mu\nu}\Sigma_{+-}^{\rho\sigma}\bigr]$ from (\ref{SErenorm}) by replacing the Lorentz interval $\Delta x^2$ with $\Delta x^2_{++}$ and $\Delta x^2_{+-}$. We will also drop the delta function terms in the $+-$ case and introduce an overall minus sign. Using (\ref{F2renorm2}) we find that,
\begin{eqnarray}
\bigl[{}^{\mu\nu}\Sigma_{\text{R}}^{\rho\sigma}\bigr](x;x') & \equiv & \kappa^2 \bigl[{}^{\mu\nu}\Sigma_{\text{R}}^{\rho\sigma}\bigr]^{(1)}(x;x') \nonumber\\
& = & \kappa^2 \mathcal{F}^{\mu\nu\rho\sigma} \Bigl[-\frac{\ln(a)}{80\pi^2} 
\delta^4(x;x') + iF_{2,\text{R}}^{(1)}(x;x')\Bigr] \; , \label{SKSE2}
\end{eqnarray}
with the S-K retarded structure function for the nonlocal part,
\begin{eqnarray}
iF_{2,\text{R}}^{(1)}(x;x') 
= \frac{i\partial^4}{5120\pi^4} 
\biggl\{ \lefteqn{ \ln^2\left(\mu^2 \Delta x^2_{++}\right) 
- \ln^2\left(\mu^2 \Delta x^2_{+-}\right)} \nonumber\\
& & \hspace{0.5cm}
-2\Bigl[ \ln\left(\mu^2 \Delta x^2_{++}\right)
- \ln\left(\mu^2 \Delta x^2_{+-}\right) \Bigr] \biggr\} \; . \label{SKF21}
\end{eqnarray}
Now, using the identities,
\begin{eqnarray}
\ln\left(\mu^2 \Delta x^2_{++}\right) 
- \ln\left(\mu^2 \Delta x^2_{+-}\right) & = & 2i\pi \theta\left(\Delta\eta - \Delta x\right) \; , \label{SKid1} \\
\ln^2\left(\mu^2 \Delta x^2_{++}\right) 
- \ln^2\left(\mu^2 \Delta x^2_{+-}\right) & = & 4i\pi \theta\left(\Delta\eta - \Delta x \right) \ln \Bigl[\mu^2\left(\Delta\eta^2 - \Delta x^2\right)\Bigr] \; , \label{SKid2}
\end{eqnarray}
we can achieve a manifestly real and causal form for $iF_{2,\text{R}}^{(1)}$,
\begin{equation}
iF_{2,\text{R}}^{(1)}(x;x') = \frac{-\partial^4}{1280\pi^3} \biggl\{
\theta\left(\Delta\eta - \Delta x \right)
\Bigl[ \ln \bigl[ \mu^2\left(\Delta\eta^2 - \Delta x^2\right) \bigr] -1 \Bigr] \biggr\}\; . \label{SKF22}
\end{equation}
Here and henceforth we define conformal time intervals $\Delta\eta \equiv \eta-\eta'$, spatial intervals $\Delta\vec{x} \equiv \vec{x}-\vec{x}'$, and its Euclidean norm $\Delta x \equiv \Vert\Delta \vec{x}\Vert$. And $\theta(\ldots)$ denotes the Heaviside step function.

\subsection{Perturbative formulation}\label{PertForm}
To get the correction to force law of gravity, we need to solve the linearized effective field equation (\ref{lineffeq}). However, we have only solved the lowest loop result for the graviton self-energy (\ref{SKSE2}),
\begin{equation}
\bigl[{}^{\mu\nu}\Sigma_{\text{R}}^{\rho\sigma}\bigr](x;x') = 0+\kappa^2 \bigl[{}^{\mu\nu}\Sigma_{\text{R}}^{\rho\sigma}\bigr]^{(1)}(x;x') + O(\kappa^4) \; . \label{SEseries}
\end{equation}
Because the quantum corrections are only known to a finite order in $\kappa^2$, there is no alternative to making a similar expansion for the gravitational potentials,
\begin{equation}
h_{\rho\sigma}(x) = h_{\rho\sigma}^{(0)}(x) + \kappa^2 h_{\rho\sigma}^{(1)}(x) + O(\kappa^4) \; . \label{hseries}
\end{equation}
Substituting (\ref{SEseries}) and (\ref{hseries}) into (\ref{lineffeq}) leads to the zeroth and first order equations,
\begin{eqnarray}
\mathcal{D}^{\mu\nu\rho\sigma}\kappa h_{\rho\sigma}^{(0)}(x) & = & \mathcal{T}_{\text{lin}}^{\mu\nu}(x) \; , \label{0theqn} \\
\mathcal{D}^{\mu\nu\rho\sigma}\kappa h_{\rho\sigma}^{(1)}(x) & = & \mathcal{S}^{\mu\nu}(x)  \; . \label{1steqn}
\end{eqnarray}
where we have the stress tensor density for the quantum corrections as
\begin{equation}
\mathcal{S}^{\mu\nu}(x) \equiv \int d^4x'\bigl[{}^{\mu\nu}\Sigma_{\text{R}}^{\rho\sigma}\bigr]^{(1)}(x;x')\kappa h_{\rho\sigma}^{(0)}(x') \; . \label{qTmn}
\end{equation}
Note that we have regarded the matter source as zeroth order, unless its stress tensor includes loop corrections from the 1PI one-point function. And the Lichnerowicz operator in de Sitter space is given by
\begin{eqnarray}
\lefteqn{
\mathcal{D}^{\mu\nu\rho\sigma} = \frac12 a^2 \Bigl[ \left(\eta^{\mu(\rho}\eta^{\sigma)\nu} - \eta^{\mu\nu}\eta^{\rho\sigma}\right)\partial^2 + \eta^{\mu\nu}\partial^{\rho}\partial^{\sigma} +\eta^{\rho\sigma}\partial^{\mu}\partial^{\nu} - 2\partial^{(\mu}\eta^{\nu)(\rho}\partial^{\sigma)}  \Bigr]
} \nonumber \\
& & \hspace{-1cm}
+Ha^3\Bigl[ \left( \eta^{\mu\nu}\eta^{\rho\sigma} - \eta^{\mu(\rho}\eta^{\sigma)\nu} \right)\partial_0 - 2\eta^{\mu\nu}\delta_0^{(\rho}\partial^{\sigma)} + 2\delta_0^{(\rho}\eta^{\sigma)(\mu}\partial^{\nu)} \Bigr] +3H^2a^4 \eta^{\mu\nu}\delta_0^{\rho}\delta_0^{\sigma} \; . \label{Lich}
\end{eqnarray}
The force law of gravity is determined from the response of a point mass. So, consider a stationary point particle with mass $M$ on a de Sitter background. We can write the matter action as
\begin{eqnarray}
S_{\text{matter}} & \equiv & -M \int d\tau \sqrt{-g_{\mu\nu}(q(\tau))\dot{q}^{\mu}(\tau)\dot{q}^{\nu}(\tau)} \nonumber\\
& = & -M \int d\tau \bigg\{a(q(\tau))\sqrt{1-\kappa h_{00}(q(\tau))}\biggr\}  \; . \label{Smatter}
\end{eqnarray}
To get from the first line to the second line in (\ref{Smatter}) we have used the full metric (\ref{FullMetric}) and have assumed the point particle has the world line $q^{\mu} \equiv \tau \delta_0^{\mu}$. So, the corresponding linearized stress tensor density takes the form
\begin{equation}
\mathcal{T}_{\text{lin}}^{\mu\nu}(x) \equiv -\kappa 
\frac{S_{\text{matter}}[h]}{\delta h_{\mu\nu}(x)}\Biggl\vert_{h = 0} = 
-8 \pi G M \delta_0^{\mu}\delta_0^{\nu} \!\times\! a \delta^3\left(\vec{x}\right)\; . \label{linstress}
\end{equation}
With this matter source, the solution to the zeroth order equation (\ref{0theqn}) is given by \cite{Tsamis:1992xa}
\begin{equation}
\kappa h^{(0)}_{\rho\sigma} = \left(2\delta_{\rho}^0\delta_{\sigma}^0 + \eta_{\rho\sigma}\right) \frac{2GM}{a\Vert\vec{x}\Vert}  \; . \label{0theqnsol}
\end{equation}
Then, integrating $\bigl[{^{\mu\nu}\Sigma_{\text{R}}^{\rho\sigma}}\bigr]^{(1)}$ up against this tree order solution $h^{(0)}_{\rho\sigma}$ (\ref{0theqnsol}) simply gives the one-loop source to the first order equation (\ref{1steqn}) which is to be solved.

\subsection{The one-loop source term}\label{Source}
The quantum source (\ref{qTmn}) consists of both local and nonlocal parts,
\begin{equation}
\mathcal{S}^{\mu\nu}(x) = \mathcal{S}_{\text{local}}^{\mu\nu}(x) + \mathcal{S}_{\text{nonlocal}}^{\mu\nu}(x) \; .
\end{equation}
The local part can be very simply expressed in terms of the linearized Weyl tensor
$C^{\mu\alpha\nu\beta}(x)$ formed from $\kappa h^{(0)}_{\rho\sigma}$ 
\cite{Leonard:2014zua},
\begin{equation}
\mathcal{S}_{\text{local}}^{\mu\nu}(x) = \partial_{\alpha} \partial_{\beta}
\Bigl[ \frac{\ln(a)}{40 \pi^2} C^{\mu\alpha\nu\beta}(x)\Bigr] \; , \label{1stsourcelocal}
\end{equation}
The nonlocal part of the source,
\begin{equation}
\mathcal{S}_{\text{nonlocal}}^{\mu\nu}(x) \equiv \mathcal{F}^{\mu\nu\rho\sigma} \int d^4x' \Bigl[iF_{2,\text{R}}^{(1)}(x;x')\Bigr]\kappa h_{\rho\sigma}^{(0)}(x')  \; , \label{1stsource}
\end{equation}
requires two steps: performing the integral and acting the derivatives.

\paragraph{\textbullet\; Fundamental integral}
\ \\
\ \\
Substituting (\ref{SKF22}) and (\ref{0theqnsol}) into (\ref{1stsource}) implies that there is a fundamental integral to do,
\begin{equation}
F(\eta,r) \equiv \int d^4x' \Biggl\{\frac{\theta(\Delta\eta-\Delta x)}{a'\Vert\vec{x}'\Vert} \biggl[ \ln\Bigl[\mu^2\left(\Delta\eta^2 - \Delta x^2\right)\Bigr]-1 \biggr]\Biggr\}  \; . \label{funint1}
\end{equation}
After changing the spatial integration variable, $\vec{x}'\rightarrow\vec{r}\equiv \Delta\vec{x}$, letting $x \equiv \Vert\vec{x}\Vert$, and performing the angular integration under spherical polar coordinates, one sees that the integral (\ref{funint1}) becomes
\begin{eqnarray}
\frac{2\pi}{x} \int_0^{+\infty} 
\lefteqn{
\left(x+r-\vert x-r\vert\right)rdr
} \nonumber \\
& & \hspace{0cm}
\times \int_{\eta_i}^{0^-} \theta(\Delta\eta - r)\biggl[\ln\Bigl[\mu^2\left(\Delta\eta^2 -r^2\right)\Bigr] - 1\biggr](-H\eta')d\eta'  \; , \label{funint2}
\end{eqnarray}
where $\eta_i\equiv -H^{-1}$ denotes the initial time. Now, let us first tackle the temporal integral. After using the Heaviside step function to fix the temporal integration limits, we find
\begin{eqnarray}
\lefteqn{
\int_{\eta_i}^{0^-} \theta(\Delta\eta - r)\biggl[\ln\Bigl[\mu^2\left(\Delta\eta^2 -r^2\right)\Bigr] - 1\biggr](-H\eta')d\eta' } \nonumber \\
& &\hspace{0.5cm}
= \int_{\eta_i}^{\eta-r} \biggl[\ln\Bigl[\mu^2\left(\Delta\eta^2 -r^2\right)\Bigr] - 1\biggr](-H\eta')d\eta'  \nonumber\\
& &\hspace{0.5cm}
= 2H(\Delta\eta_i+\eta_i)r\ln(2\mu r)
+ H(2\Delta\eta_i+3\eta_i-r)(\Delta\eta_i-r) \nonumber\\
& &\hspace{1cm}
-\frac12 H(\Delta\eta_i+2\eta_i+r)(\Delta\eta_i+r)\ln\bigl[\mu(\Delta\eta_i+r)\bigr] \nonumber\\
& &\hspace{1.5cm}
-\frac12 H(\Delta\eta_i+2\eta_i-r)(\Delta\eta_i-r)\ln\bigl[\mu(\Delta\eta_i-r)\bigr] \; , \label{tint}
\end{eqnarray}
where we have defined $\Delta\eta_i \equiv \eta-\eta_i$. Note that the result (\ref{tint}) will force the integration limit of the radial integral in (\ref{funint2}) to be $0<r<\Delta\eta_i$. Next step is to substitute (\ref{tint}) into (\ref{funint2}), and we are left with a purely radial integration which is tedious but straightforward to do. Here, we simply give the full result of the fundamental integral (\ref{funint1}),
\begin{eqnarray}
\lefteqn{
F(\eta,x) = \frac{2\pi H}{3x}\times \biggl\{
-(\Delta\eta_i+\eta_i)x^4\ln(2\mu x) } \nonumber \\
& &\hspace{-0.5cm}
+\frac1{600}\Bigl[ 5(119\Delta\eta_i+440\eta_i)\Delta\eta_i^3x + 1250(\Delta\eta_i+\eta_i)x^4 - 10(131\Delta\eta_i+220\eta_i)\Delta\eta_ix^3 - 261x^5 \Bigr]
  \nonumber\\
& &\hspace{0cm}
+\frac1{20}\Bigl[ (\Delta\eta_i+10\eta_i)x - 2(\Delta\eta_i+5\eta_i)\Delta\eta_i + 3x^2\Bigr](\Delta\eta_i+x)^3 \ln\bigl[\mu(\Delta\eta_i+x)\bigr]
 \nonumber\\
& &\hspace{0.5cm}
+\frac1{20}\Bigl[ (\Delta\eta_i+10\eta_i)x + 2(\Delta\eta_i+5\eta_i)\Delta\eta_i - 3x^2\Bigr](\Delta\eta_i-x)^3 \ln\bigl[\mu(\Delta\eta_i-x)\bigr]  \biggl\} \; . \label{funintsol}
\end{eqnarray}

\paragraph{\textbullet\; Acting derivatives}
\ \\
\ \\
At this stage, we can rewrite the nontrivial part of our nonlocal source term (\ref{1stsource}) as
\begin{equation}
\mathcal{S}_{\text{nonlocal}}^{\mu\nu}(\eta,x) = -\frac{GM}{640\pi^3}\left(2\delta_{\rho}^0 \delta_{\sigma}^0 + \eta_{\rho\sigma}\right) \mathcal{F}^{\mu\nu\rho\sigma} \partial^4 F(\eta,x)  \; . \label{1stsource2}
\end{equation}
Next is to act two d'Alembertians on $F(\eta,x)$ using the following differentiation formula:
\begin{equation}
\partial^4 F(\eta,x) = F^{(4,0)} -2F^{(2,2)} + F^{(0,4)} -\frac4{x}F^{(2,1)} + \frac4{x} F^{(0,3)} \; , \label{del4F}
\end{equation}
where we have adopted a shorthand notation of mixed partial differentiations,
\begin{equation}
F^{(m,n)} \equiv \frac{\partial^{m+n}}{\partial \eta^m\partial x^n}F(\eta,x) \; . \label{mixpar}
\end{equation}
Then, it follows that two d'Alembertians on (\ref{funintsol}) gives
\begin{equation}
G(\eta,x) \equiv \partial^4 F(\eta,x) = 16\pi H \left[1-\frac{\eta}{x}\ln\left(2\mu x\right)\right] \; . \label{del4Fsol}
\end{equation}
Now the remaining work is acting the operator $\left(2\delta_{\rho}^0 \delta_{\sigma}^0 + \eta_{\rho\sigma}\right) \mathcal{F}^{\mu\nu\rho\sigma}$ on $G(\eta,x)$. Note that the tensor projector $\mathcal{F}^{\mu\nu\rho\sigma}$ is traceless by design; therefore, what we really care about is the result of acting $\left(2\delta_{\rho}^0 \delta_{\sigma}^0\right) \mathcal{F}^{\mu\nu\rho\sigma}$,
\begin{equation}
2\delta_{\rho}^0 \delta_{\sigma}^0 \mathcal{F}^{\mu\nu\rho\sigma} 
= \partial^{\mu}\partial^{\nu}\partial_0^2 
- \frac13\partial^{\mu}\partial^{\nu}\overline{\partial}^2
- 2\delta_0^{(\mu}\partial^{\nu)}\partial^2\partial_0
+ \frac13\eta^{\mu\nu}\partial^2\partial_0^2
+ \left(\! \delta_0^{\mu}\delta_0^{\nu} + \frac13\eta^{\mu\nu}\! \right)\partial^4 \; . \label{00Fproj}
\end{equation}
Similarly, we can also derive a formula for this differential operation,
\begin{eqnarray}
\lefteqn{
2\delta_{\rho}^0 \delta_{\sigma}^0 \mathcal{F}^{\mu\nu\rho\sigma} G(\eta,x) = \overline{\partial}^{\mu}\overline{\partial}^{\nu}\left(G^{(2,0)} - \frac13 G^{(0,2)} - \frac2{3x} G^{(0,1)}\right) } \nonumber\\
& & \hspace{0.5cm}
+ \delta_0^{(\mu}\overline{\partial}^{\nu)}\left(-\frac43 G^{(1,2)} - \frac8{3x} G^{(1,1)} \right)
+ \delta_0^{\mu}\delta_0^{\nu}\left(\frac23 G^{(0,4)} + \frac8{3x} G^{(0,3)} \right)  \nonumber\\
& & \hspace{1cm}
+ \overline{\eta}^{\mu\nu}\left(-\frac13 G^{(2,2)} +\frac13 G^{(0,4)} - \frac2{3x} G^{(2,1)} +\frac4{3x} G^{(0,3)} \right) \; . \label{00FprojG}
\end{eqnarray}
Substituting the result of $G(\eta,x)$ (\ref{del4Fsol}) into (\ref{00FprojG}) allows us to finally obtain the nonlocal part of our one-loop source term,
\begin{equation}
\mathcal{S}_{\text{nonlocal}}^{\mu\nu}(\eta,x) = \frac{GM}{40\pi^2} \Biggl\{
\overline{\partial}^{\mu}\overline{\partial}^{\nu}\!\left(\frac{H\eta}{3x^3}\right) 
- \delta_0^{\mu}\delta_0^{\nu}\! \left(\frac{4H\eta}{x^5} \right)
- \overline{\eta}^{\mu\nu}\! \left(\frac{2H\eta}{x^5}\right) 
+ \delta_0^{(\mu}\overline{\partial}^{\nu)}\!
\left(\frac{4H}{3x^3}\right) \!\Biggr\} \; . \label{1stsource3}
\end{equation}
Substituting (\ref{1stsourcelocal}) and (\ref{1stsource3}) into (\ref{1steqn}), and ignoring
the delta function contributions, gives the following nontrivial components of our first order equation:
\begin{eqnarray}
\mathcal{D}^{00\rho\sigma}\kappa h_{\rho\sigma}^{(1)}(\eta,\vec{x}) & = & 
\frac{GM}{10\pi^2 a x^5} \; , \label{001steqn} \\
\mathcal{D}^{0i\rho\sigma}\kappa h_{\rho\sigma}^{(1)}\bigl(\eta,\vec{x}\bigr) & = & \partial^i \biggl\{\frac{GMH}{60\pi^2 x^3} \biggr\} \; , \label{0i1steqn}\\
\mathcal{D}^{ij\rho\sigma}\kappa h_{\rho\sigma}^{(1)}\bigl(\eta,\vec{x}\bigr) & = & \partial^i\partial^j\biggl\{-\frac{GM}{120\pi^2 a x^3} \!+\!
\frac{GM H^2}{40\pi^2 a x} \biggr\} 
+ \delta^{ij}\biggl\{\frac{GM}{20\pi^2 a x^5} \biggr\} \; . \label{ij1steqn}
\end{eqnarray}

\subsection{Quantum corrected potentials}\label{LoopSol}
We want to express the one-loop Newtonian potentials in Schwarzschild coordinates so that with a convenient gauge choice their nonzero components take the form
\begin{equation}
\kappa h_{00}^{(1)}\bigl(\eta,\vec{x}\bigr) = f_1\bigl(\eta,\vec{x}\bigr) \;\;\; , \;\;\; \kappa h_{0i}^{(1)}\bigl(\eta,\vec{x}\bigr) = 0 \;\;\; , \;\;\; \kappa h_{ij}^{(1)}\bigl(\eta,\vec{x}\bigr) = f_2\bigl(\eta,\vec{x}\bigr)\delta_{ij} \; . \label{ansatz}
\end{equation}
Substituting our ansatz (\ref{ansatz}) into (\ref{001steqn})-(\ref{ij1steqn}) gives rise to the following equations
\begin{eqnarray}
\frac{GM}{60\pi^2 a^3 x^3} \!-\! \frac{GM H^2}{20 \pi^2 a x}
& = & f_1 - f_2 \; , \label{heqn1} \\
\frac{GM}{60\pi^2 a^3 x^3}
& = & \eta f_2^{(1,0)} - f_1 \; , \label{heqn2} \\
\frac{GM}{10\pi^2 a^3 x^5}
& = & f_2^{(0,2)} + \frac2{x}f_2^{(0,1)} + \frac3{\eta}f_2^{(1,0)} - \frac3{\eta^2}f_1 \; , \label{heqn3} \\
\frac{GM}{20\pi^2 a^3 x^5}
& = & \lefteqn{ \frac12 f_1^{(0,2)} + \frac1{x} f_1^{(0,1)} - \frac1{\eta} f_1^{(1,0)} + \frac3{\eta^2} f_1 } \nonumber \\
& & \hspace{0cm}
-\frac12 f_2^{(0,2)} - \frac1{x}f_2^{(0,1)} - \frac2{\eta} f_2^{(1,0)} + f_2^{(2,0)} \;. \label{heqn4} 
\end{eqnarray}
We should remind readers of the notation for mixed partial differentiation (\ref{mixpar}). Now it seems that we have four equations (\ref{heqn1})-(\ref{heqn4}) but only two field variables, $f_1$ and $f_2$. Actually, these equations are not completely independent. We can pick the easier ones to solve; then, the remaining equations shall be automatically satisfied by the virtue of stress energy tensor conservation and the Bianchi identity. 

Equations (\ref{heqn1}) and (\ref{heqn2}) are much simpler to start with, so we first eliminate $f_1$ using these two and obtain a first order partial differential equation for $f_2$,
\begin{equation}
\partial_0 f_2 + H a f_2 = -\frac{GMH}{30\pi^2 a^2 x^3} + \frac{G M H^3}{20 \pi^2 a} \; , \label{f2eqn}
\end{equation}
whose solution is
\begin{equation}
f_2\bigl(\eta,\vec{x}\bigr) = \frac{GM}{60\pi^2 a^3 x^3} 
+ \frac{G M H^2 \ln(a)}{20 \pi^2 a x} + \frac{C(x)}{a} \; . \label{f2solnot}
\end{equation}
Note that $C(x)$ is a integration ``constant,'' but it may still depend on the Euclidean norm $x$. To fix this constant, we can substitute (\ref{heqn1}) and (\ref{f2solnot}) into (\ref{heqn3}) and obtain a second order ordinary differential equation for $C(x)$,
\begin{equation}
C''(x) + \frac2{x} C'(x) = -\frac{GM H^2}{20\pi^2 x^3} \; , \label{Creqn}
\end{equation}
which can be regarded as first order in $C'(x)$ and hence easily solved. Thus, we find
\begin{equation}
C(x) = \frac{GM H^2}{20\pi^2 x} \times \Bigl[1+\ln\left(Hx\right)\Bigr] +\frac{A}{x} + B \; , \label{Crsol}
\end{equation}
where $A$ and $B$ are really constants. To fix them we shall, in principle, make use of (\ref{heqn4}). Well, as we have mentioned previously, whichever equation is left should be satisfied automatically. And indeed, with the solutions given by (\ref{heqn1}) and (\ref{f2solnot}) and (\ref{Crsol}), one can show that (\ref{heqn4}) is solved no matter what $A$ and $B$ are. Therefore, we can simply set $A=B=0$, and we obtain the formal solutions for $f_1$ and $f_2$,
\begin{eqnarray}
f_1\bigl(\eta,\vec{x}\bigr) & = & \frac{GM}{30\pi^2 a^3 x^3}
+\frac{GM H^2 \ln(a H x)}{20\pi^2 a x} \; , \label{h100sol} \\
f_2\bigl(\eta,\vec{x}\bigr) & = & \frac{GM}{60\pi^2 a^3 x^3}
+\frac{GM H^2 [1 \!+\! \ln(aHx)]}{20\pi^2 a x} \; . \label{h1ijsol}
\end{eqnarray}
Combining the tree order solutions (\ref{0theqnsol}) gives the final results of our one-loop Newtonian potentials,
\begin{eqnarray}
\kappa^2 \Bigl[\kappa h_{00}^{(1)}\bigl(\eta,\vec{x}\bigr)\Bigr]\!\!\!\! & = & \!\!\!\!\Biggl\{\!\frac{\kappa^2}{60\pi^2 (ax)^2} \!+\! \frac{\kappa^2 H^2 \ln(aHx)}{40\pi^2} \!+\! O(\kappa^4) \!\Biggr\}\!\! \times \!\! \Bigl[\kappa h_{00}^{(0)}\bigl(\eta,\vec{x}\bigr)\Bigr] \; , \label{h00sol} \\
\kappa^2 \Bigl[\kappa h_{0i}^{(1)}\bigl(\eta,\vec{x}\bigr)\Bigr]\!\!\!\! & = & \!\!O(\kappa^4)\; , \label{h0isol} \\
\kappa^2 \Bigl[\kappa h_{ij}^{(1)}\bigl(\eta,\vec{x}\bigr)\Bigr]\!\!\!\! & = & \!\!\!\!\Biggl\{\!\frac{\kappa^2}{120\pi^2 (ax)^2} \!+\! \frac{\kappa^2 H^2 [1 \!+\! \ln(aHx)]}{40\pi^2}
\!+\! O(\kappa^4) \!\Biggr\}\!\! \times \!\! \Bigl[\kappa h_{ij}^{(0)}\bigl(\eta,\vec{x}\bigr)\Bigr] \; . \label{hijsol}
\end{eqnarray}

\section{DISCUSSION}\label{discuss}

We have computed the one photon loop contribution to the graviton self-energy on 
de Sitter space, and we have used it to derive quantum corrections to the gravitational potentials 
of a stationary point mass. Our results (\ref{h00sol})-(\ref{hijsol}) feature two sorts
of terms which can best be understood by expressing the full Newtonian potential
as its classical value times a series of corrections, with the factors of $\hbar$ 
and $c$ restored,
\begin{equation}
\Phi(\eta,\vec{x}) = -\frac{2GM}{a x} \Biggl\{1 + \frac{4 \hbar G}{15 \pi c^3 (a x)^2} 
+ \frac{2 \hbar G H^2}{5 \pi c^5} \Bigl[ \ln\Bigl(\frac{a H x}{c}\Bigr) + 1\Bigr] + 
O(G^2) \Biggr\} . \label{Phi}
\end{equation}
The fractional correction $4 \hbar G/[15 \pi c^3 (a x)^2]$ is just the de Sitter 
version of the flat space background correction first found in 1970 by Radkowski 
\cite{AFR} and later confirmed by Capper, Duff and Halpern \cite{Capper:1974ed}. 
Getting this term right checks an important correspondence limit; however, it can 
only be significant at physical distances of $a x \sim \sqrt{\hbar G/c^3} \sim 
10^{-35}~$m. This is a small distance even with respect to the scales prevailing 
during primordial inflation. The more interesting part of (\ref{Phi}) is the 
fractional correction $2 \hbar G H^2/[5 \pi c^5] \times \ln(aHx/c)$, which vanishes 
in the flat space limit. This represents a small logarithmic running of the Newton 
constant. One interesting consequence is that an observer carried away by the 
Hubble flow at constant $x$ sees the one loop correction to the gravitational
potential grow with respect to its classical value, although classical and
quantum potentials both fall off exponentially in co-moving time. Hence a very 
long phase of inflation leads to a breakdown of perturbation theory.

The logarithmic running correction in (\ref{Phi}) might seem surprising because 
it is sourced by the gravitational vacuum polarization induced by photons, which
is simply related to its value on flat space by the conformal invariance of 
classical electromagnetism. Indeed, the source terms we found in Eqs 
(\ref{001steqn})-(\ref{ij1steqn}) are almost identical to those of flat space. 
The sources on the right-hand side of (\ref{001steqn}) and (\ref{ij1steqn}) are
just conformal rescalings of the flat space result, and the new, ``de Sitter''
source on the right-hand side of (\ref{0i1steqn}) is the bare minimum required by
conservation in view of the time dependent conformal rescaling. The reason this
conformally rescaled source can do more than ``de Sitterize'' the response of the flat
background is that the left-hand side of the equation is different because 
gravity is not conformally invariant. Gravity in a de Sitter background responds
far differently to sources than gravity does in a flat background.  

It is worthwhile working out the response seen by a static observer whose background
geometry is
\begin{equation}
ds^2 = -\left(1-H^2r^2\right)dt^2 + \frac{dr^2}{(1-H^2r^2)} + r^2 d\Omega^2 \; .
\end{equation}
The transformation to these coordinates is
\begin{eqnarray}
\eta = \frac{-H^{-1} e^{-Ht}}{\sqrt{1 \!-\! H^2 r^2}} & \Longrightarrow &
d\eta = \frac1{a} \Biggl[ dt - \frac{ Hr dr}{1 \!-\! H^2 r^2} \Biggr] 
\; , \label{trans1} \\
x^i = \frac{r^i e^{-Ht}}{\sqrt{1 \!-\! H^2 r^2}} & \Longrightarrow &
dx^i = \frac1{a} \Biggl[ -H r^i dt + dr^i + \frac{H^2 r^i r dr}{1 \!-\! H^2 r^2}
\Biggr] \; . \label{trans2}
\end{eqnarray} 
The potentials come from using transformations (\ref{trans1})-(\ref{trans2}) on
the full metric $g_{\mu\nu} \equiv a^2 [\eta_{\mu\nu} + \kappa h_{\mu\nu}]$,
\begin{eqnarray}
ds^2_{\rm full} & \equiv & a^2 \Bigl[-d\eta^2 + d\vec{x} \!\cdot\! d\vec{x}\Bigr] 
+ \kappa h_{00} a^2 d\eta^2 + 2 \kappa h_{0i} a^2 d\eta dx^i + \kappa h_{ij} a^2
dx^i dx^j \; , \\
& \equiv & -\Bigl[1 \!-\! H^2 r^2\Bigr] dt^2 \!+\! \frac{dr^2}{1 \!-\! H^2 r^2} 
+ r^2 d\Omega^2 \!+\! \chi_{tt} dt^2 \!+\! 2 \chi_{0i} dt dr^i \!+\! \chi_{ij} 
dr^i dr^j \; . \qquad 
\end{eqnarray}
The final results are complicated by the fact that achieving the de Sitter-Schwarzschild 
form would require order $GM$ modifications of the
transformations (\ref{trans1}) and (\ref{trans2}). However, one important feature of
the result is that all of the factors of $a x$ become the time independent static 
coordinate radius $r$. Hence an observer at constant physical distance from the source
detects only a temporally constant, logarithmic enhancement of the potential. 

\hspace{1cm} 

\centerline{\bf Acknowledgements}

We are grateful for our conversations and correspondence on this subject 
with M. Fr\"ob, S. Park, T. Prokopec, and E. Verdaguer. In particular, Fr\"ob 
and Verdaguer pointed out a mistake in our treatment of the local contribution
to the source term which resulted in our original results possessing a factor of
$\ln(Hx)$ without the corresponding factor of $\ln(a)$ that appears in their
solution \cite{Frob:2016fcr}, and in the similar result for a massless, minimally
coupled scalar \cite{Park:2015kua}. This work was partially supported by 
NSF Grant No. PHY-1506513 and by the Institute for Fundamental Theory at the 
University of Florida.

\end{document}